\documentclass[usenatbib]{article}
\pdfoutput=1
\usepackage[latin9]{inputenc}
\usepackage[a4paper]{geometry}
\usepackage{color}
\usepackage{float}
\usepackage{graphicx}

\makeatletter

\providecommand{\tabularnewline}{\\}

\usepackage{jcappub}

\usepackage{astrobib_mnras2e}

\makeatother

\begin{document}

\title{Efficient Construction of Mock Catalogs for Baryon Acoustic Oscillation Surveys}

\author{Tomomi Sunayama\textsuperscript{a}, Nikhil Padmanabhan\textsuperscript{a},
Katrin Heitmann\textsuperscript{b}, Salman Habib\textsuperscript{b},
Esteban Rangel\textsuperscript{b,c}}

\abstract{Precision measurements of the large scale structure of the
  Universe require large numbers of high fidelity mock catalogs to
  accurately assess, and account for, the presence of systematic
  effects. We introduce and test a scheme for generating mock catalogs
  rapidly using suitably derated N-body simulations. Our aim is to
  reproduce the large scale structure and the gross properties of dark
  matter halos with high accuracy, while sacrificing the details of
  the halo's internal structure.  By adjusting global and local
  time-steps in an N-body code, we demonstrate that we recover halo
  masses to better than $0.5\%$ and the power spectrum to better than $1\%$ both in real and
  redshift space for $k=1h{\rm Mpc^{-1}}$,
  while requiring a factor of 4 less CPU time.  We also calibrate the
  redshift spacing of outputs required to generate simulated
  light cones. We find that outputs separated by $\Delta z=0.05$ allow
  us to interpolate particle positions and velocities to reproduce the
  real and redshift space power spectra to better than $1\%$ (out to
  $k=1h{\rm Mpc^{-1}}$).  We apply these ideas to generate a suite of
  simulations spanning a range of cosmologies, motivated by the Baryon
  Oscillation Spectroscopic Survey (BOSS) but broadly applicable to
  future large scale structure surveys including eBOSS and DESI. As an
  initial demonstration of the utility of such simulations, we
  calibrate the shift in the baryonic acoustic oscillation peak
  position as a function of galaxy bias with higher precision than has
  been possible so far. This paper also serves to document the
  simulations, which we make publicly available.}

\affiliation{\textsuperscript{a} Department of Physics, Yale
  University, New Haven, CT 06511}

\affiliation{\textsuperscript{b} High Energy Physics and Mathematics
  \& Computer Science Divisions, Argonne National Laboratory, Lemont,
  IL 60439}

\affiliation{\textsuperscript{c} Department of Electrical Engineering
  and Computer Science, Northwestern University, Evanston, IL 606208}

\emailAdd{tomomi.sunayama@yale.edu, nikhil.padmanabhan@yale.edu,heitmann@anl.gov,
habib@anl.gov,steverangel@u.northwestern.edu }


\maketitle

\section{Introduction}

Large-volume spectroscopic surveys of the Universe
\citep{2013arXiv1305.5422S,2000AJ....120.1579Y,2011MNRAS.415.2876B}
are revolutionizing our understanding of cosmology and structure
formation. Based on these successes, a new generation of surveys
\citep{2013arXiv1308.0847L,2013arXiv1305.5422S,2011arXiv1110.3193L} is
being planned that will improve cosmological constraints by an order
of magnitude (or more). This unprecedented increase in statistical
precision places stringent demands on the underlying theoretical
modeling and analysis techniques; numerical simulations will play an
essential role in meeting these requirements.

One set of challenges for simulations arise from the varied roles they
play, and the different requirements these impose in turn on the
simulations. At one extreme, simulations are necessary for estimating
the errors on the measurements.  This typically requires very large
volumes to simulate entire surveys thousands of times, but have lower
accuracy requirements. Motivated by these considerations, a number of
recent studies have investigated methods designed to produce mock
catalogs with reduced accuracy, but much higher throughput compared to
full N-body
simulations~\cite{2002ApJ...564....8M,2002MNRAS.331..587M,2008MNRAS.391..435F,
  2013AN....334..691R,2013arXiv1312.2013C,2013JCAP...06..036T,2014MNRAS.437.2594W,
  2013MNRAS.433.2389M,2001A&A...367...18H,2009ApJ...701..945S,2014MNRAS.439L..21K,2014arXiv1409.1124C}.
The effect of changing the input cosmology used to generate the
covariance matrix on cosmological inferences is still not fully
understood, nor is it clear how best to implement such variations (but
see Refs.~\citep{Morrison2013} and \citep{Kalus2015} for recent work
on this topic). More recently, the impact of super-survey modes (modes
outside the survey volume) on inferred errors has been shown to be
potentially larger than previously appreciated
(e.g., Ref.~\cite{2013PhRvD..87l3504T}) and is an area of active study.

At the other extreme, simulations are crucial for calibrating the
theoretical models used to fit the data. Examples here are quantifying
shifts in the baryon acoustic oscillation distance scale due to
nonlinear evolution and galaxy bias
\citep{2008ApJ...686...13S,2008PhRvD..77b3533C,2008PhRvD..77d3525S,2009PhRvD..80f3508P,2010ApJ...720.1650S,2012PhRvD..85j3523S}, 
or templates used to fit the full shape of the galaxy correlation
function.  For such applications, one ideally requires high fidelity
simulations.  The volume requirements are significantly reduced from
that for covariance matrices, but the simulations still need to cover volumes
much larger than survey volumes to keep systematic errors below
statistical errors.

An intermediate application is the generation of mock catalogs that
capture the observational characteristics of surveys (e.g., geometry,
selection effects). The importance of these cannot be underestimated,
since the effects of many observational systematics can only be
quantitatively estimated by simulating them. These issues will become
progressively more important for the next generations of surveys which
will move away from highly complete and pure samples that have mostly
been used for cosmological studies to date.

The simplest way to generate approximate density fields is to use
analytic approximations such as Lagrangian perturbation theory
followed by prescriptions that place halos within the density field
to match the halo distribution measured in N-body
simulations~\cite{2013MNRAS.428.1036M,2014arXiv1401.4171M}, or simply
to run lower resolution N-body codes with a small number of
time-steps~\cite{2013JCAP...06..036T}, or a combination of the two
approaches~\cite{2014MNRAS.437.2594W}. These methods are successful in
capturing the large-scale density field but lose information on small
scales. Because of their speed, they can be used to produce large
numbers of simulations required to build sample covariance matrices,
at error levels ranging from $5-10\%$ (depending on the quantities
being predicted). It is difficult to estimate, however, what the loss
of accuracy implies for tests of systematic errors, which may need to
be modeled at the $< 1\%$ level.

The approach we take here is to reduce the small spatial scale accuracy of a
high-resolution N-body code by coarsening its temporal resolution.
For the code we consider, the time-stepping consists of two
components, (i) a long time step for solving for the evolution under
the long-range particle-mesh (PM) force, and (ii) a set of underlying
sub-cycled time steps for a short-range particle-particle interaction,
computed either via a tree-based algorithm, or by direct
particle-particle force evaluations. The idea is to reduce the number
of both types of time steps while preserving enough accuracy to
correctly describe the large scale distribution of galaxies, as
modeled by a halo occupation distribution (HOD) approach. One goal in
this paper is, therefore, to quantitatively understand the impact of
the temporal resolution on the halo density field and how to
accurately reproduce the details of the halo density field on large
scales, while sacrificing small scale structure information. Doing this
successfully allows us to generate a suite of large volume
simulations. 

Having determined the optimal time-stepping that maintains accuracy on the scales of interest, we then generate
a suite of simulations using this procedure. We use these 
simulations to demonstrate two example applications discussed above.
The first example addresses the generation of mock catalogs for spectroscopic 
surveys. We focus on the Baryon Oscillation Spectroscopic Survey
(BOSS), although these simulations can be used to simulate 
aspects of future surveys such as eBOSS and DESI. The gains from optimizing
our time-stepping allow us to simulate a volume large enough to embed 
two entire BOSS volumes into the same simulation. As part of 
building these mock catalogs, we also consider the frequency with 
which we need to store snapshots of the N-body simulation in order
to include light-cone effects in the mocks.

Our second application highlights the usefulness of these techniques
for constraining systematic effects: we use the simulations 
to constrain shifts in the BAO acoustic scale as a function of 
halo bias. This requires very large volumes to statistically 
measure the shifts, while at the same time, one must robustly determine 
the halo properties as well. Although we restrict ourselves here to a 
simple demonstration, our results show an unambiguous shift in the 
BAO scale due to a combination of nonlinear evolution of the density
field, as well as halo bias.

The paper is organized as follows. Section~\ref{sec:hacc} briefly
describes the Hardware/Hybrid Accelerated Cosmology Code (HACC) N-body
framework we use to generate our simulations, focusing on the
flexibility in the time-stepping that we exploit
here. Section~\ref{sec:time} presents a sequence of convergence tests
where we evaluate the effects of time-stepping on the halo density
field. Section~\ref{sec:light cones} discusses interpolation between
saved time steps, necessary for constructing light cone
outputs. Section~\ref{sec:boss} presents the two example applications
of the simulations; we conclude in Section~\ref{sec:disc} by outlining
possible future directions.

All simulations and calculations in this
paper assume a $\Lambda$CDM cosmology with $\Omega_{m}=0.2648$,
$\Omega_{\Lambda}=0.7352$, $\Omega_{b}h^{2}=0.02258$, $n_{s}=0.963$,
$\sigma_{8}=0.8$ and $h=0.71$.

\section{Coarse-Graining in Time with HACC}
\label{sec:hacc}

All simulations in this paper were carried out using the Hardware/Hybrid Accelerated Cosmology Code (HACC)
framework. HACC provides an advanced, architecture-agile,
extreme-scale N-body capability targeted to cosmological
simulations. In this Section, we describe how HACC's hierarchical time-stepping scheme can be easily used to implement coarse-graining in time so as to reduce the CPU time but maintaining acceptable accuracy for halo statistics.

\subsection{The HACC Framework}

HACC is descended from an approach originally developed for the heterogeneous architecture of Roadrunner
\cite{1742-6596-180-1-012019,Pope:2010:AU:1845737.1845828}, the first supercomputer to break the petaflop performance barrier. HACC's flexible code architecture combines MPI with a variety of more
local programming models, (e.g., OpenCL, OpenMP) and is easily
adaptable to different platforms. HACC has demonstrated scaling on the
entire IBM BG/Q Sequoia system up to 1,572,864 cores with an equal
number of MPI ranks, attaining 13.94 PFlops at 69.2\% of peak and 90\%
parallel efficiency (for details, see Ref.~\citep{2012arXiv1211.4864H}). Examples of science results obtained using HACC include 64-billion particle runs for baryon acoustic oscillations predictions for the BOSS Lyman-$\alpha$ forest \cite{2010ApJ...713..383W}, high-statistics predictions for the halo profiles of massive clusters \cite{2013ApJ...766...32B}, and 0.55 and 1.1~trillion particle runs at high mass resolution~\cite{Heitmann2014}. A recent overview of the HACC
framework can be found in Ref.~\citep{hacc_2014}.

HACC uses a hybrid parallel algorithmic structure, splitting the force
calculation into a specially designed grid-based long/medium range
spectral particle mesh (PM) component that is common to all computer
architectures, and an architecture-specific short-range
solver. Modular code design combined with particle caching allows the
short-range solvers to be `hot-swappable' on-node; they are blind to
the parallel implementation of the long-range solver. The short-range
solvers can use direct particle-particle interactions, i.e., a
P$^{3}$M algorithm \cite{1988csup.book.....H}, as on (Cell or GPU)
accelerated systems, or use tree methods on conventional or many-core
architectures. (This was the case for the simulations reported here.)
In all cases, the time-stepping scheme is based on a symplectic method
with (adaptive) sub-cycling of the short-range force. The availability
of multiple algorithms within the HACC framework allows us to carry
out careful error analyses, for example, the P$^{3}$M and the TreePM
versions agree to within $0.1\%$ for the nonlinear power spectrum test
in the code comparison suite of Ref.~\cite{2005ApJS..160...28H}.

\subsection{HACC Time Stepping Scheme}

As already discussed, an important feature of the work presented here
is the ability to carry out error-controlled approximate simulations
at high throughput. In order to understand how to implement this, we
provide some details on the HACC time-stepping algorithm. In HACC, time  
evolution is viewed as a symplectic map on phase space: $\zeta(t)=\exp(-t{\bf  {H}})\zeta(0)$ where, $\zeta$ is a phase-space vector $({\bf x},{\bf
  v})$, $H$ is the (self-consistent) Hamiltonian, and the operator,
${\bf {H}}=[H,~]_{P}$, denotes the action of taking the Poisson
bracket with the Hamiltonian.  Suppose that the Hamiltonian can be
written as the sum of two parts; then by using the
Campbell-Baker-Hausdorff (CBH) series we can build an integrator for
the time evolution; repeated application of the CBH formula yields
\[
\exp[-t({\bf {H}}_{1}+{\bf {H}}_{2})]=\exp[-(t/2){\bf
  {H}}_{1}]\exp(-t{\bf {H}}_{2})\exp[-(t/2){\bf {H}}_{1}]+O(t^{3}), 
\]
a second order symplectic integrator. In the basic PM application, the
Hamiltonian $H_{1}$ is the free particle (kinetic) piece while $H_{2}$
is the one-particle effective potential; corresponding respectively to
the `stream' and `kick' maps $M_{1}=\exp(-t{\bf {H}}_{1})$ and
$M_{2}=\exp(-t{\bf {H}}_{2})$. In the stream map, the particle
position is drifted using its known velocity, which remains unchanged;
in the kick map, the velocity is updated using the force evaluation,
while the position remains unchanged. This symmetric `split-operator'
step is termed SKS (stream-kick-stream). A KSK scheme constitutes an
alternative second-order symplectic integrator.

In the presence of both short and long-range forces, we split the
Hamiltonian into two parts, $H_{1}=H_{sr}+H_{lr}$ where $H_{sr}$
contains the kinetic and particle-particle force interaction (with an
associated map $M_{sr}$), whereas, $H_{2}=H_{lr}$ is just the long
range force, corresponding to the map $M_{lr}$. Since the long range
force varies relatively slowly, we construct a single time-step map by
sub-cycling $M_{sr}$:
$M_{full}(t)=M_{lr}(t/2)[M_{sr}(t/n_{c})]^{n_{c}}M_{lr}(t/2)$, the
total map being a usual second-order symplectic integrator. This
corresponds to a KSK step, where the S is not an exact stream step,
but has enough $M_{sr}$ steps composed together to obtain the required
accuracy. (We take care that the time-dependence in the
self-consistent potential is treated correctly; HACC uses the scale
factor, $a$, as the time variable.)  The code therefore has two
degrees of freedom to tune its time-steps: the length of the full time step ($t$ above), and the number of sub-cycles for the short range force ($n_c$
above).  As discussed later below, we will use the flexibility in the
sub-cycling as a way of reducing the number of time steps such that
the loss of accuracy only affects the resolution at very small scales,
which, as discussed previously, are not of interest in the current set
of simulations. For high-resolution applications, HACC allows for particle-level adaptive time-stepping, but that feature is not exploited here.

\section{Time Step Tuning and Halo Matching}
\label{sec:time}

In this section, we systematically examine how reducing the number of global
time steps affects individual gross halo properties (i.e., halo
masses, positions, and velocities), as well as aggregate statistics
such as the halo mass function and spatial clustering. We run a set of
convergence tests with boxes of size $(256h^{-1}{\rm Mpc})^{3}$ with
$256^{3}$ particles. These runs have a similar particle mass as the
full $(4000h^{-1}{\rm Mpc})^{3}$ volume simulations we present
later. We run these with the following time step options : 450/5,
300/3, 300/2, 150/3 and 150/2 where the first number is the number of
long time-steps, while the second is the number of subcycles.  The
450/5 case has been independently verified to give fully converged
results and is the baseline against which we compare all other
results.  Each simulation is started from the same initial condition
and evolved down to $z=0.15$. We demonstrate that the 300/2 case,
corresponding to a time step of $\Delta a\approx0.003$ reproduces the
full resolution simulation for all the large scale properties we
consider, and is our choice for the mock catalogs presented in
Section~\ref{sec:boss}.

\subsection{Halo Matching Algorithm}

In order to compare detailed halo properties, we need to match individual
halos between our reference 450/5 run and test runs with reduced time steps. 
All simulations share the same particle initial conditions, allowing
us to match halos in different runs by matching their individual
particle content. Given a halo in simulation A, we consider all halos
in simulation B that between them hold all the particles belonging to
the halo in simulation A. Given this list of possible matches, we
choose the halo in run B with the largest number of common particles
with the reference halo in run A. To avoid spurious matches, we also
require that the fraction of common particles (relative to simulation
A) exceeds a given threshold. To illustrate how this matching
algorithm works, we use the samples from the 300/2 simulation and the
450/5 simulation, and adopt a threshold of 50\% as our default choice. Our results 
are insensitive to the exact choice of the this threshold.

\begin{figure}[h]
\includegraphics[width=0.465\columnwidth] {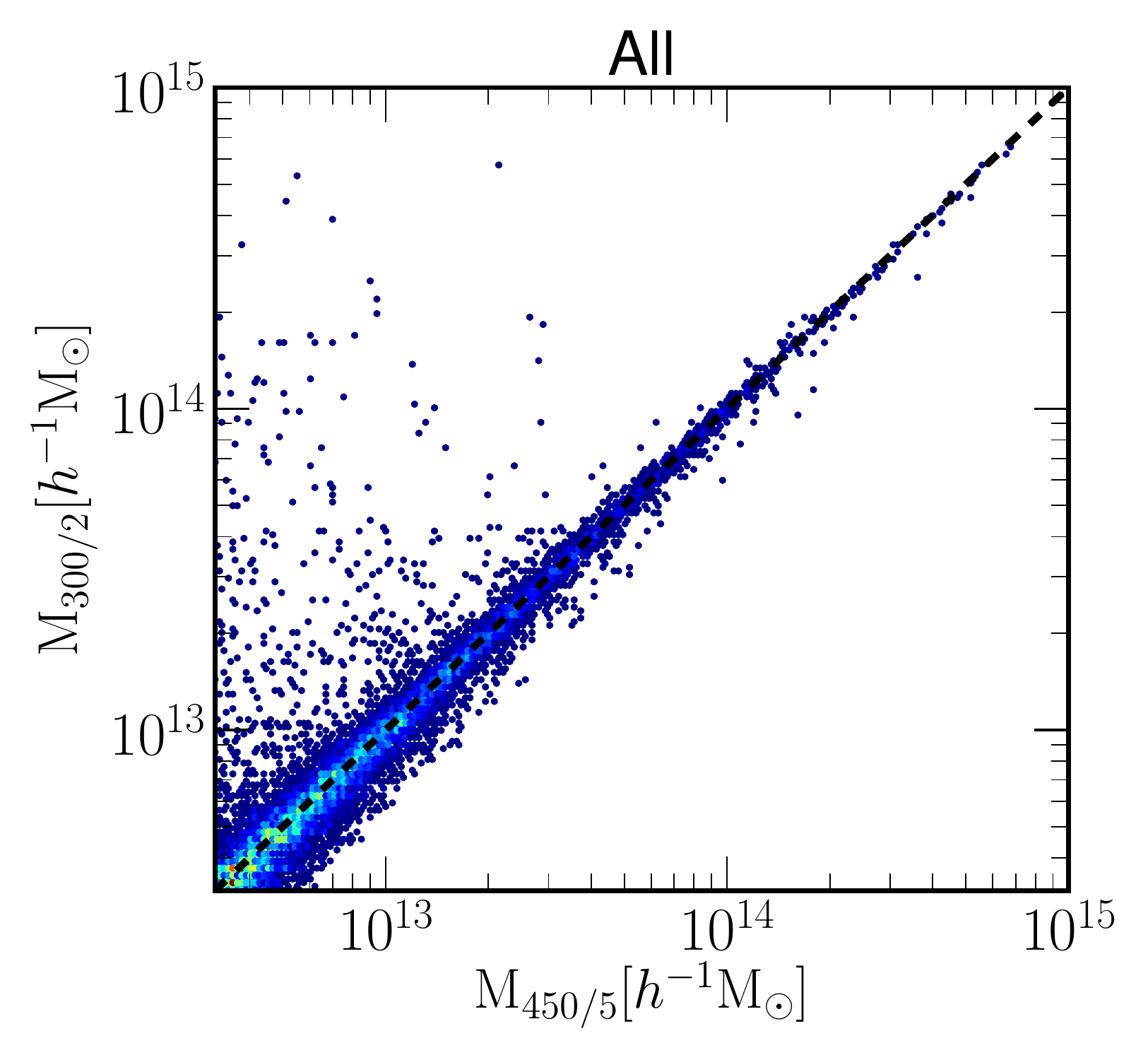} 
\includegraphics[width=0.535\columnwidth] {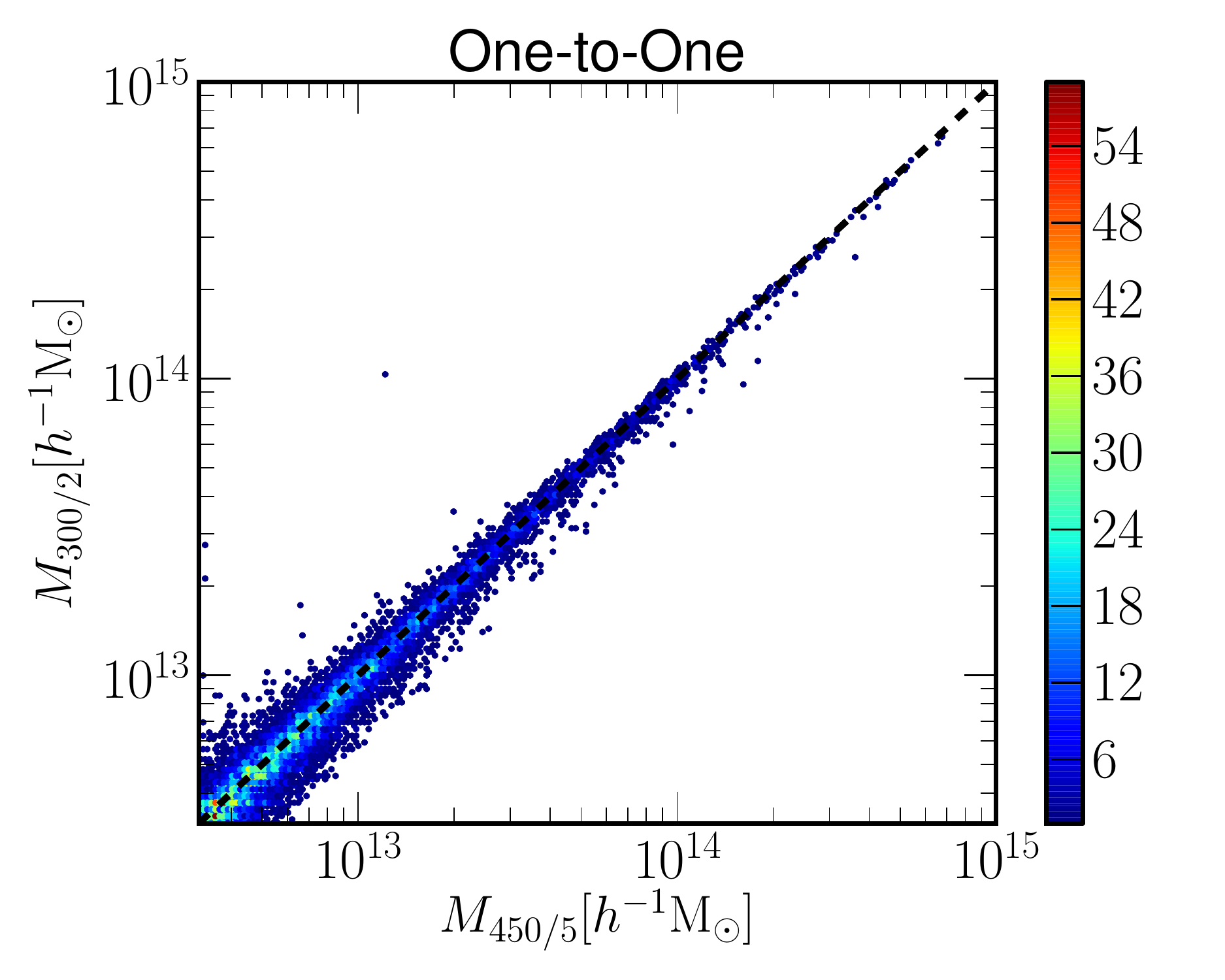} 

\includegraphics[width=0.465\textwidth] 
{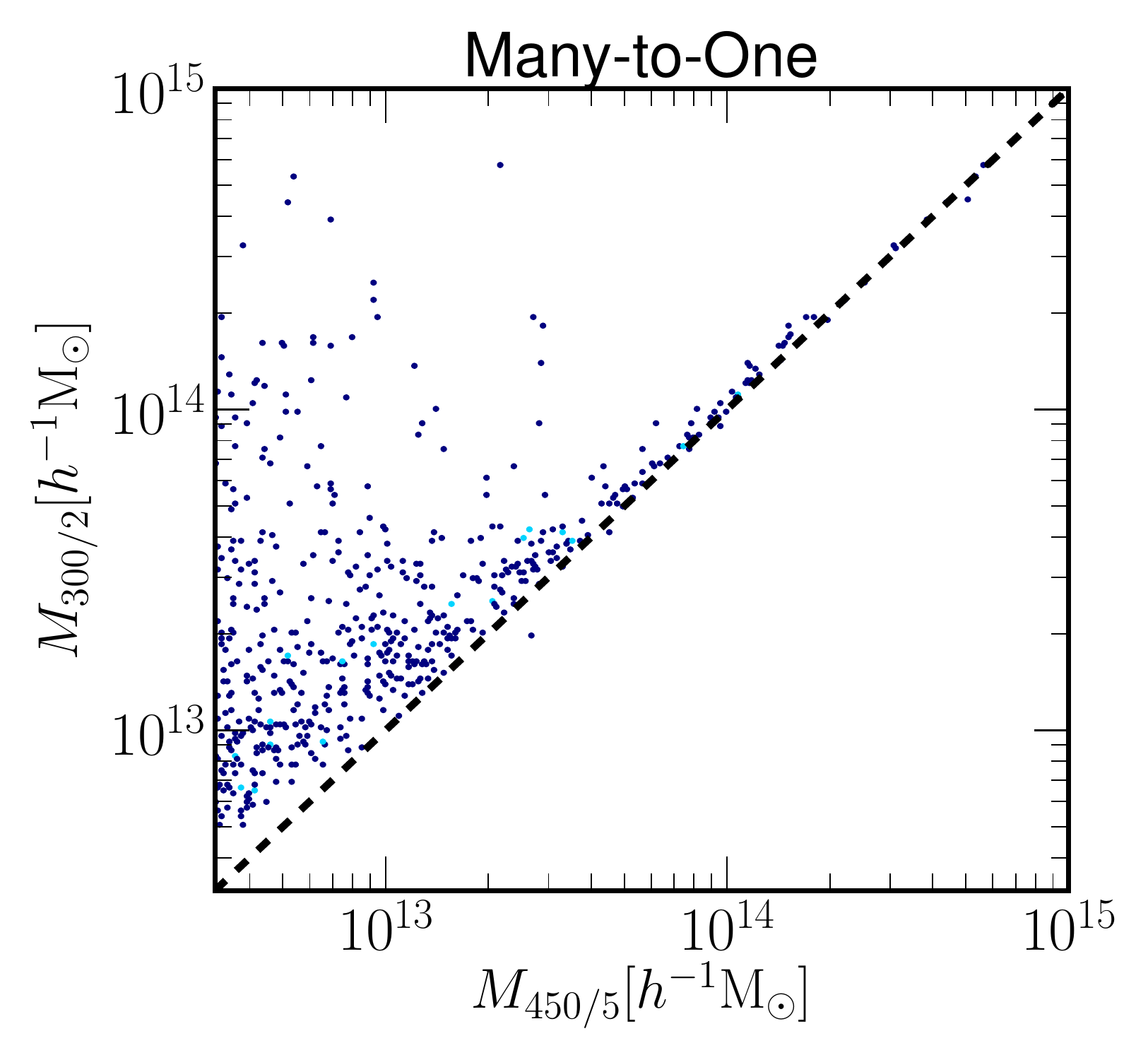} 
\includegraphics[width=0.535\columnwidth] 
{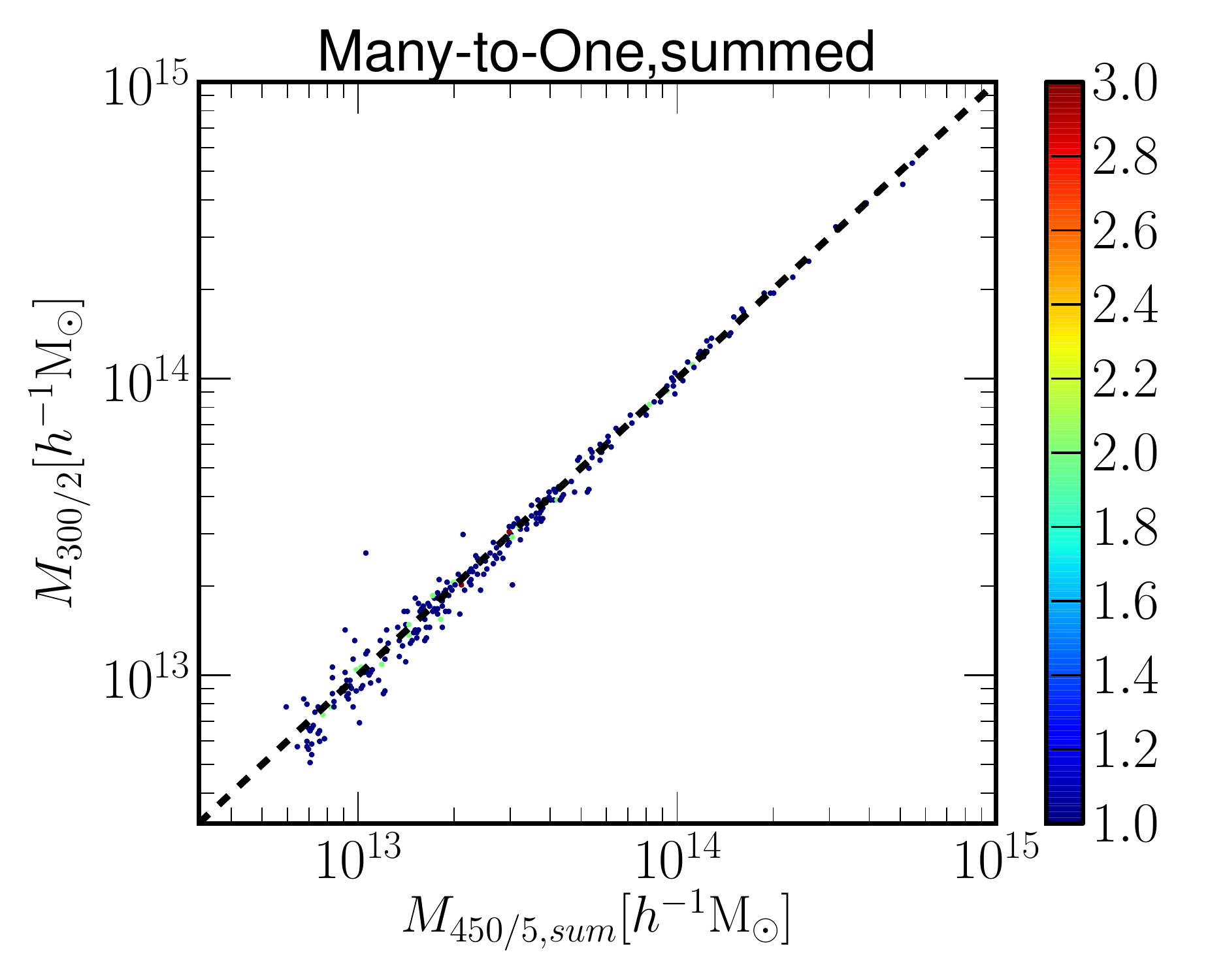} 

\caption{\label{fig:mass-scatter1}A two-dimensional histogram comparing the 
halo mass of matched halos in the 450/5 simulation (x-axis) to the 300/2 simulation
  (y-axis) at $z=0.15$. Panels correspond to halos with different
  matching criteria imposed: all the matched halos (top left), the
  vast majority of matched halos having one-to-one correspondence (top
  right), matched halos not having one-to-one correspondence called
  ``multiply-booked'' halos (bottom left), and the multiply-booked
  halos whose corresponding halo masses are added (bottom right). The colors 
  denote the number of halos in each bin. The results shown in these panels imply that the low-mass scatter
  between the 450/5 simulation and the 300/2 simulation shown in the
  top left panel arises when physically associated halos in the 450/5
  simulation are merged into one halo in the 300/2 simulation due to
  an effectively worse resolution in the latter case.}
\end{figure}

The matching algorithm described above is unidirectional -- multiple
halos in run A may have particles resident in a single halo in run
B. In our simulations, this happens at the 1-2\% level, adopting a
particle matching threshold of 50\%. We refer to these cases as
`multiply-booked' halos. Figure~\ref{fig:mass-scatter1} compares halo
masses when matching the 450/5 simulation to the 300/2 simulation for
the case of multiply-booked halos, as well as the rest. The top left
panel shows the mass scatter for all the matched halos between the two
simulations, while the top right panel shows the mass scatter only for
the halos that have one-to-one correspondence.  The bottom left
panel shows the mass scatter for individual multiply-booked halos,
while the bottom right panel plots the summed halo mass for the
corresponding halos. The overall behavior represented in
Figure~\ref{fig:mass-scatter1} is straightforward to interpret.

As the top left panel shows, there are low-mass halos in the 450/5
simulation matched to high-mass halos in the 300/2 simulation.  The
same trend is observed for the case of multiply-booked halos (bottom
left panel), but not for the one-to-one matched halos (top right).
Furthermore, the disagreement for halo masses between the two
simulations are resolved by adding the corresponding halo masses. This
implies that there are multiple halos in the 450/5 simulation which
are merged into one halo in the 300/2 simulation. The smaller number
of time steps in the 300/2 simulations reduces substructure as well as
the compactness of the halos compared to the 450/5 simulation. Thus,
for a small fraction of halos in the 450/5 simulation, individual
halos can be merged into a single halo in the 300/2 simulation.

\begin{figure}[H]
\includegraphics[width=0.5\columnwidth]{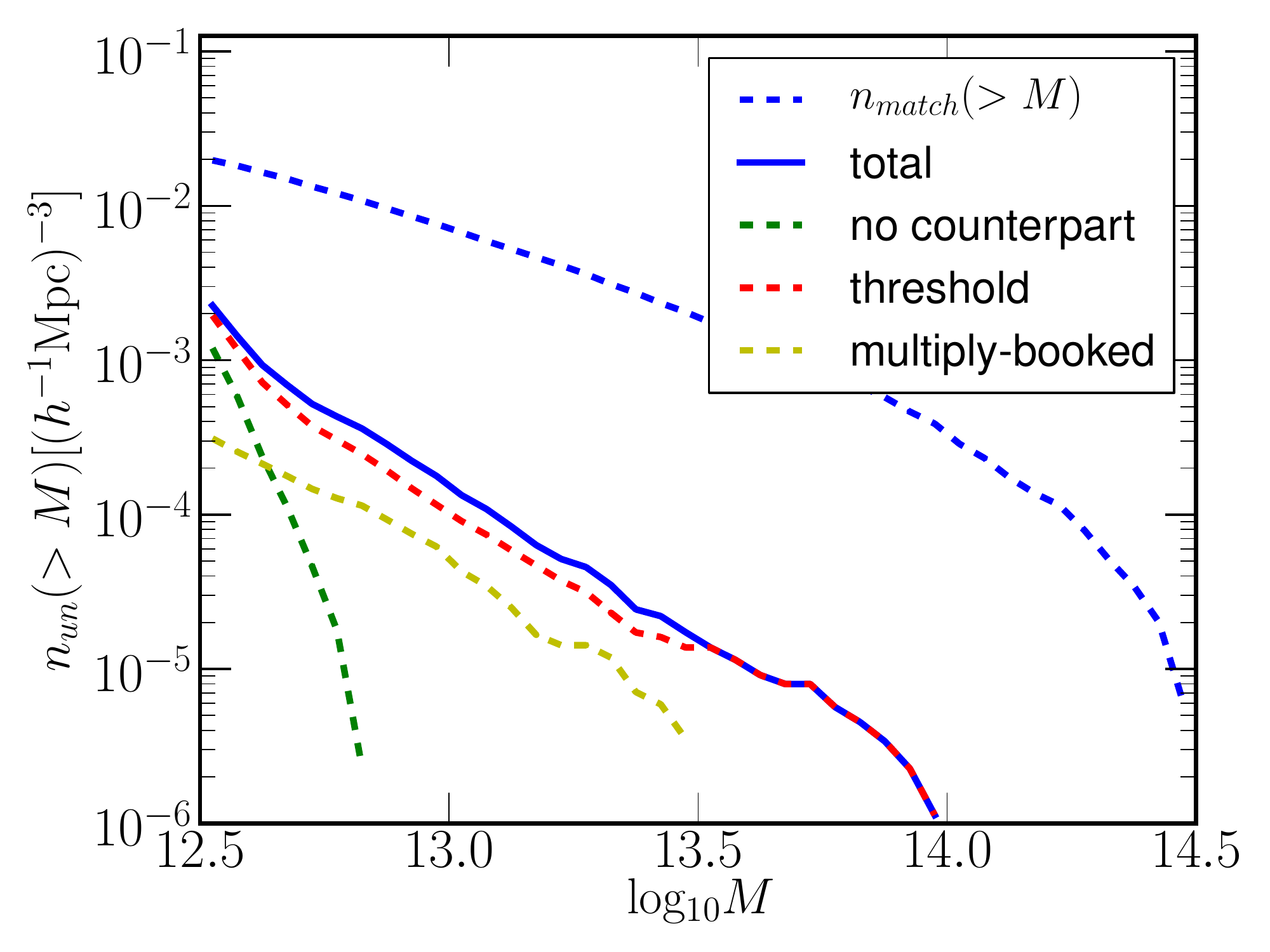} 

\caption{\label{fig:mass-content} Itemization of unmatched halos (from the
450/5 and 300/2 simulations at $z=0.15$) shown as cumulative number
densities of the unmatched halos arising from each procedure in the
matching algorithm. The solid blue line is the total number density
of the unmatched halos. The dashed green line shows halos with no
counterpart -- none of the particles were identified as belonging
to a halo in the comparison simulation; this is significant only at
low halo mass. The dashed red line shows halos eliminated because
of not meeting the matching threshold (i.e., the halos do not have
enough of a fraction of the same particles). The dashed yellow line
is for the halos eliminated because multiple halos correspond to one
halo (see text). The dashed line is the cumulative number density of matched halos, shown for reference.}
\end{figure}

Figure \ref{fig:mass-content} shows the number densities of the
unmatched halos in the 450/5 simulation when compared to the 300/2
simulation at $z=0.15$. There are three reasons that halos can turn up
as unmatched.  In the first case, particles forming a halo in
simulation A may not form a component of a halo in simulation B (no
common particles).  Second, if the fraction of common particles over
the total number of particles in each halo is less than the threshold
of 50\%, these halos will be eliminated from the matching
set. Finally, for the case of multiply-booked halos, we remove all but
the one with the largest number of common particles. In Figure
\ref{fig:mass-content}, we show each type of unmatched number density
as a function of halo mass.  The first case occurs only for low mass
halos, where low effective resolution in a simulation can lead to halo
drop out (halos are too ``fuzzy'' to meet the Friends-of-Friends (FOF)
connectivity criterion), and falls off steeply with rising halo
mass. Most of the unmatched halos arise because they do not pass the threshold criterion.  The fraction of unmatched halos due to being
``multiply-booked'' is similar to the threshold case, albeit at a
lower level.

The trends described above persist at different redshifts. As we
reduce the number of total long timesteps taken, the unmatched
fraction increases, due to the lower resolution of the simulations,
while the number of sub-cycles does not noticeably change these
results. In Section~3.2 where we compare halo properties, we restrict
ourselves to halos with a one-to-one correspondence (i.e., not
multiply-booked halos).

\subsection{Halo Properties}

We now systematically compare halo properties (i.e., halo mass,
position, and velocity) for halos in the lower resolution runs that
were successfully matched to those in the 450/5 simulation. We are
interested in correctly describing the large-scale distribution of
galaxies using an HOD approach; this requires that only the dark
matter halo locations and masses be estimated with sufficient
accuracy. 

\begin{figure}[h]
\textcolor{black}{\includegraphics[width=0.5\columnwidth]{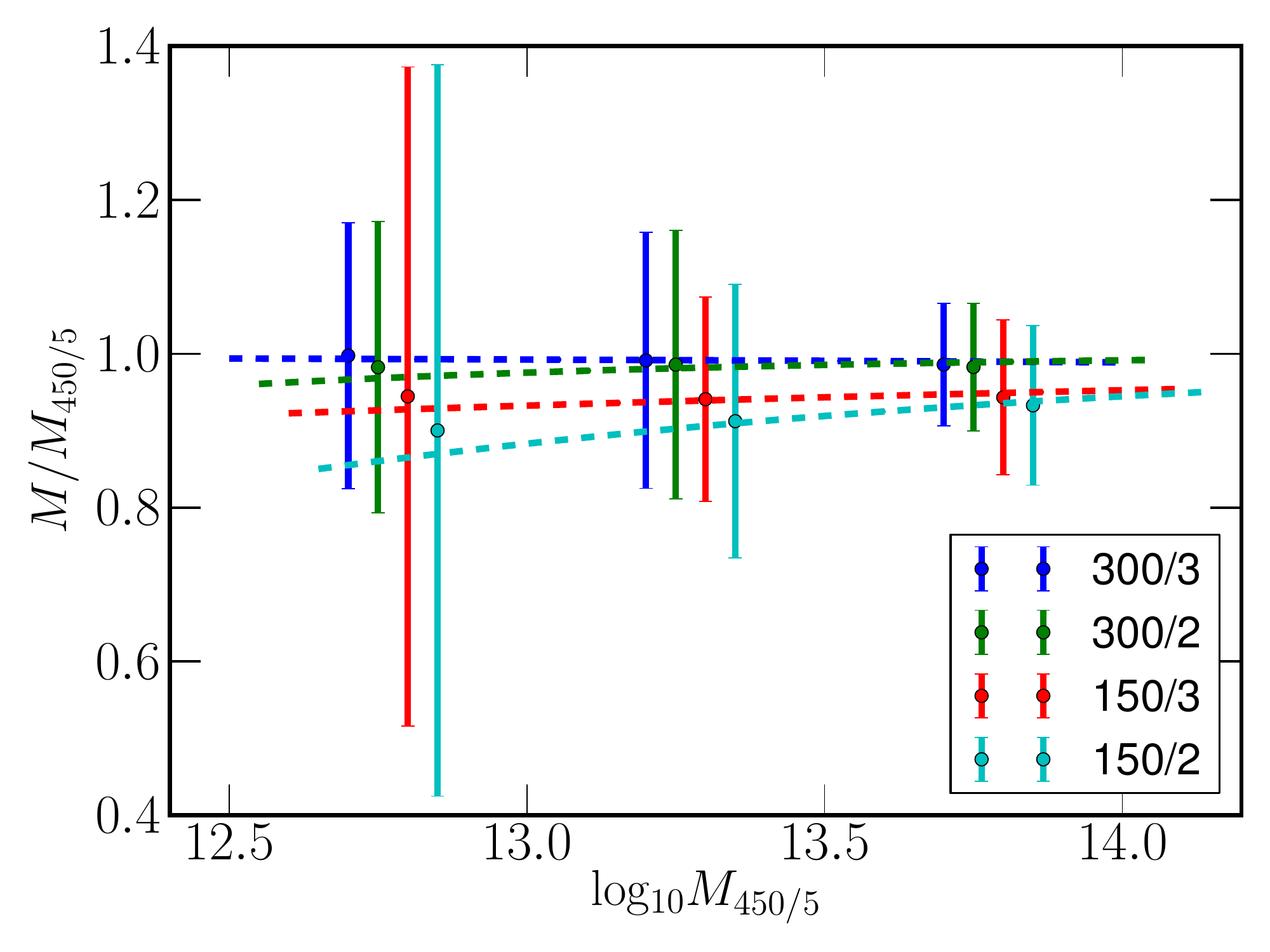}} 

\caption{\label{fig:HaloProperty_mass}Comparison of halo mass (FOF,
  $b=0.168$) for matched halos between the 450/5 simulation and
  coarsened time-stepping schemes at $z=0.15$. We take all the matched
  halos whose masses are between $10^{12.5}h^{-1}{\rm M_{\odot}}$ to
  $10^{13.0}h^{-1}{\rm M_{\odot}}$, $10^{13.0}h^{-1}{\rm M_{\odot}}$ to
  $10^{13.5}h^{-1}{\rm M_{\odot}}$, and $10^{13.5}h^{-1}{\rm M_{\odot}}$ to
  $10^{14.0}h^{-1}{\rm M_{\odot}}$, and compute the mean and the standard
  deviation for ${M/M_{450/5}}$ where $M_{450/5}$ is a halo mass
  for the 450/5 simulation and $M$ stands for the simulations with
  different number of time steps corresponding to different colors in
  the plot. The dashed lines plot the mass recalibration formula we 
  apply (Eq.~\ref{eq:mass_adjust}).The $x$-positions have been displaced to avoid overlapping
  the error bars. Halo masses decrease systematically as the time
  resolution is coarsened.  }
\end{figure}

The comparison of halo masses for different time-stepping schemes to the
450/5 simulation at $z=0.15$ is shown in Figure
\ref{fig:HaloProperty_mass}.  We take all the matched halos whose
masses are between $10^{12.5}h^{-1}{\rm M_{\odot}}$ to $10^{13.0}h^{-1}{\rm
  M_{\odot}}$, $10^{13.0}h^{-1}{\rm M_{\odot}}$ to $10^{13.5}h^{-1}{\rm
  M_{\odot}}$, and $10^{13.5}h^{-1}{\rm M_{\odot}}$ to $10^{14.0}h^{-1}{\rm
  M_{\odot}}$, and compute their means and standard deviations for
${\rm M/M_{450/5}}$, where $M_{450/5}$ is a halo mass for the 450/5
simulation and $M$ corresponds to a halo in the samples generated with
different time-stepping schemes. Figure \ref{fig:HaloProperty_mass}
shows that the simulations with small number of time steps produce
systematically lower FOF masses than the 450/5 simulation. (The
same linking length ($b=0.168$) is used in the FOF algorithm to define
halos for all simulations.) For the case of the 300/2 simulation,
the deviation from the FOF masses in the 450/5 simulation is about
1.7\%, while for the case of the 150/3 and 150/2 simulations, the
deviations are about 5.7\% and 8.5\% respectively. The FOF masses are 
highest in the 450/5 simulation,
decreasing systematically with increase in loss of temporal
resolution.

\begin{figure}[h]
  \includegraphics[width=0.5\columnwidth]{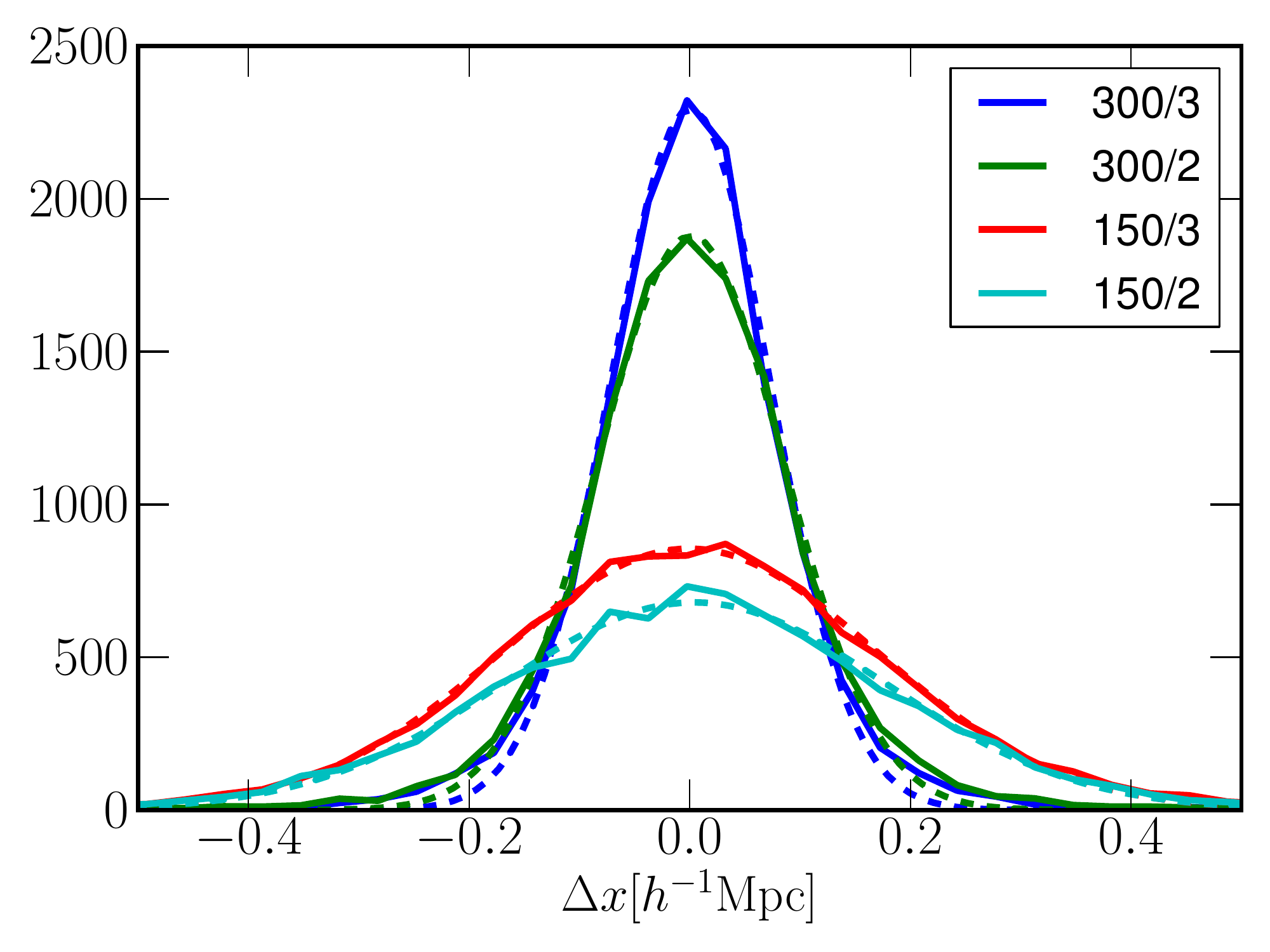}\includegraphics[width=0.5\columnwidth]{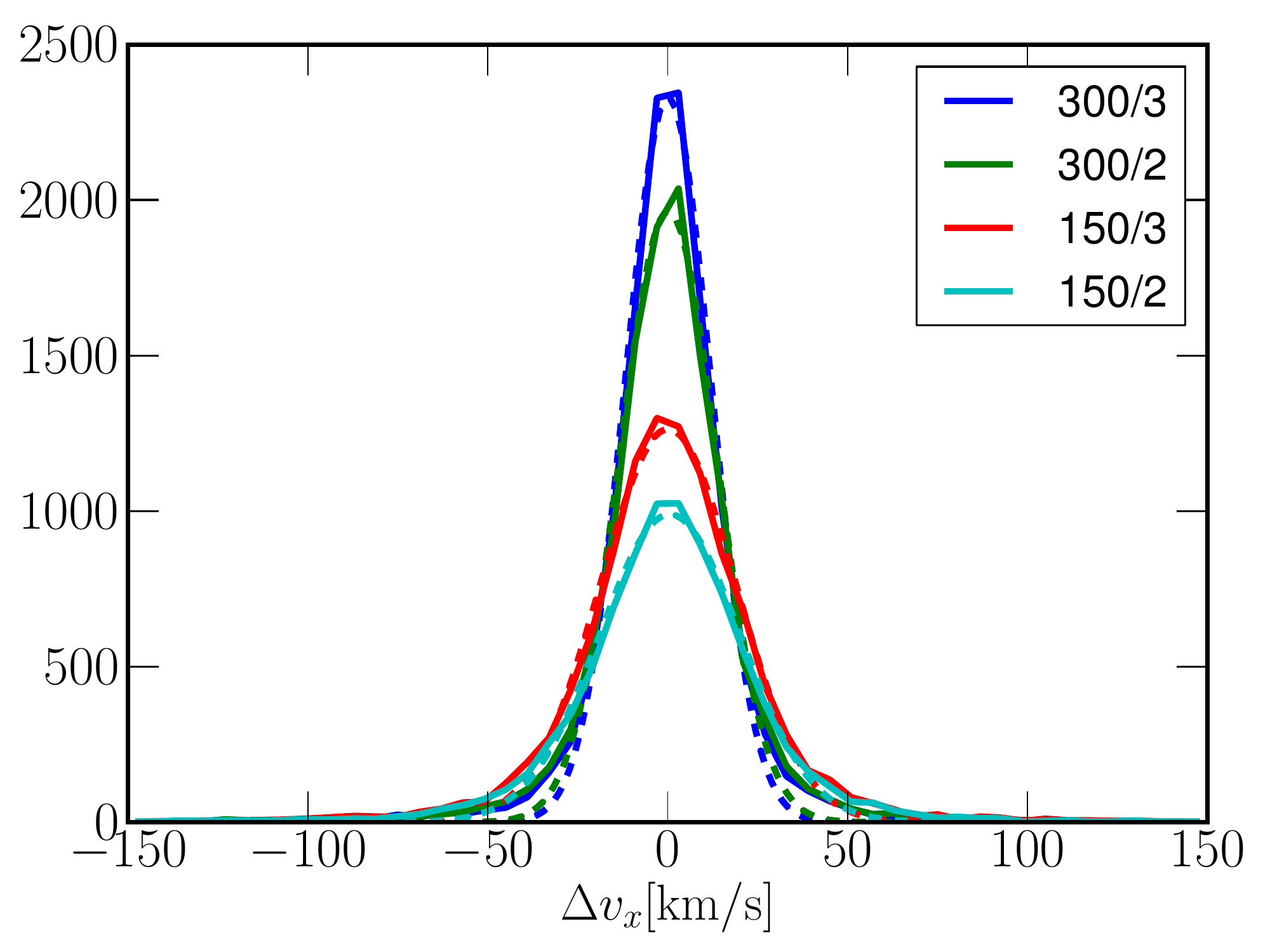}

\caption{\label{fig:HaloProperty_step}Comparison of the positions (left) and
velocities (right) of halos matched across simulations with different
time steps. The reference simulation is 450/5 while the colors correspond
to 300/3 (blue), 300/2 (green), 150/3 (red), and 150/2 (cyan). The
dashed lines are Gaussian fits.}
\end{figure}

Figure~\ref{fig:HaloProperty_step} shows the differences in positions
(left panel) and velocities (right panel) for the matched halos at
$z=0.15$. Simulations with a smaller number of global time steps (150)
display significantly more scatter; they also show a small bias in the
velocity. With 300 global time steps, the results are much improved;
the velocity bias is almost entirely removed and the scatter is
significantly reduced. The standard deviation in the differences in
halo distances is matched to better than $100h^{-1}{\rm kpc}$ in these
cases, and better than $15{\rm km/s}$ in velocities. As a 
reference, the standard deviation of velocities for the full resolution
simulation (450/5) is about $300{\rm km/s}$. The distributions are
very close to Gaussian, as shown by the dashed lines. As is
clear from Figure~\ref{fig:HaloProperty_step}, the difference between
3 and 2 sub-cycles is insignificant for our purposes. We observe the
same trends for the halo properties discussed here at different redshifts.

As shown in Figure \ref{fig:mass-content}, the fraction of unmatched
halos in the 300/2 simulation to the 450/5 simulation is less than
$2\%$ over most of the halo mass ranges and less than $1\%$ for halo masses
greater than $10^{13.0}h^{-1}{\rm M_{\odot}}$, which implies that the 300/2
simulation has almost the same number of halos as the 450/5
simulation. Furthermore, Figure \ref{fig:mass-scatter1} shows that
most of the halos in the 300/2 simulation have the same mass as the
ones in the 450/5 simulation. Since the number of sub-cycles only
affects the halo positions and velocities at a low level, as shown in
Figure \ref{fig:HaloProperty_step}, we choose the 300/2 time step to
save simulation time while keeping the halo properties almost
identical to the 450/5 simulation.

The results shown in Figures~\ref{fig:mass-scatter1},
\ref{fig:mass-content}, and \ref{fig:HaloProperty_step}, show that the
300/2 option has a low ratio of unmatched halos (less than $2\%$),
excellent halo mass correlation to the 450/5 simulation (the small
mass bias can be easily corrected as described below), and
sufficiently small scatter in halo position. This time-stepping option
is therefore a good candidate for generating mock catalogs
efficiently, while maintaining high accuracy.  In terms of the time
savings alone, this will result in an increased capacity to generate
high quality catalogs by a factor of four, which is quite
significant. We will consider memory and storage savings further below
in Section~\ref{sec:light cones}.

\subsection{Halo Mass Adjustment and Resulting Observables}

Although the halo mass is a surprisingly robust quantity as a function of time-step (see, e.g., Figure~8 of Ref.~\cite{2007ApJ...671.1160L}), halos generated by the de-tuned simulations do have systematically lower
masses than halos in the 450/5 simulation as shown in
Figure~\ref{fig:HaloProperty_mass}.  In the following, we describe how
to implement a systematic mass correction by matching to the 450/5
results; we also display the resulting observables including mass
functions and power spectra.

To undertake the mass calibration, we first take all the matched halos
between the 450/5 simulation and the de-tuned simulations and compute
means for each mass bin. We consider only the matched halos because
the aim of the mass adjustment is to correct systematic mass differences
for the halos that are identical in the different runs.
After computing the means for each mass bin, we fit them to a functional
form that brings the reassigned halo mass, $M_{re}$, close to the
average halo mass for the 450/5 simulation. For our simulations, we
find that the following simple form suffices for this task:

\begin{equation}
M_{re}=M[1.0+\alpha(M/10^{12.0}[h^{-1}{\rm M_{\odot}])^{\beta}],}
\label{eq:mass_adjust}
\end{equation}
where $M_{re}$ is the reassigned halo mass, $M$ is the original halo
mass, and $\alpha$ and $\beta$ are free parameters. The $\alpha$ and
$\beta$ values for the simulations with different numbers of time
steps are listed in Table \ref{tab:free_param1} (at $z=0.15$).  The
best-fit parameters $\alpha$ and $\beta$ are functions of redshift.
For the case of the 300/2 simulation, the best fit parameters are
$\alpha(z)=0.123z+0.052$ and $\beta(z)=-0.154z-0.447$.

\begin{table}
\begin{tabular}{|c|c|c|}
\hline 
 & $\alpha$  & $\beta$\tabularnewline
\hline 
\hline 
300/3  & 0.005  & 0.175\tabularnewline
\hline 
300/2  & 0.07  & -0.47\tabularnewline
\hline 
150/3  & 0.101  & -0.162\tabularnewline
\hline 
150/2  & 0.315  & -0.411\tabularnewline
\hline 
\end{tabular}

\caption{\label{tab:free_param1}Mass reassignment parameters $\alpha$
  and $\beta$ from Eq.~\ref{eq:mass_adjust} for simulations run with
  different numbers of time steps (the results are shown at $z=0.15$).}
\end{table}

\begin{figure}[H]
\includegraphics[width=0.5\columnwidth] {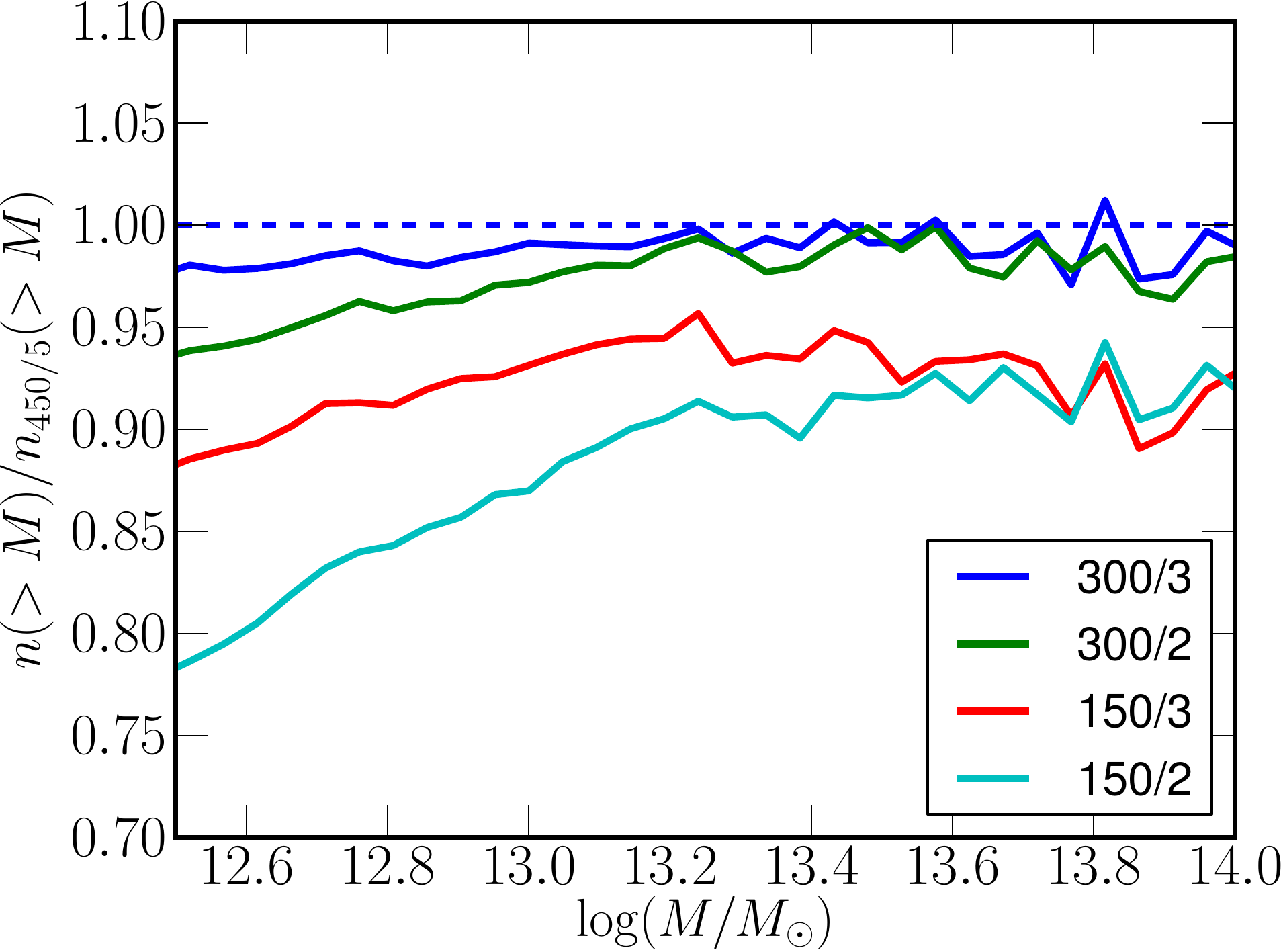} 
\includegraphics[width=0.5\columnwidth] {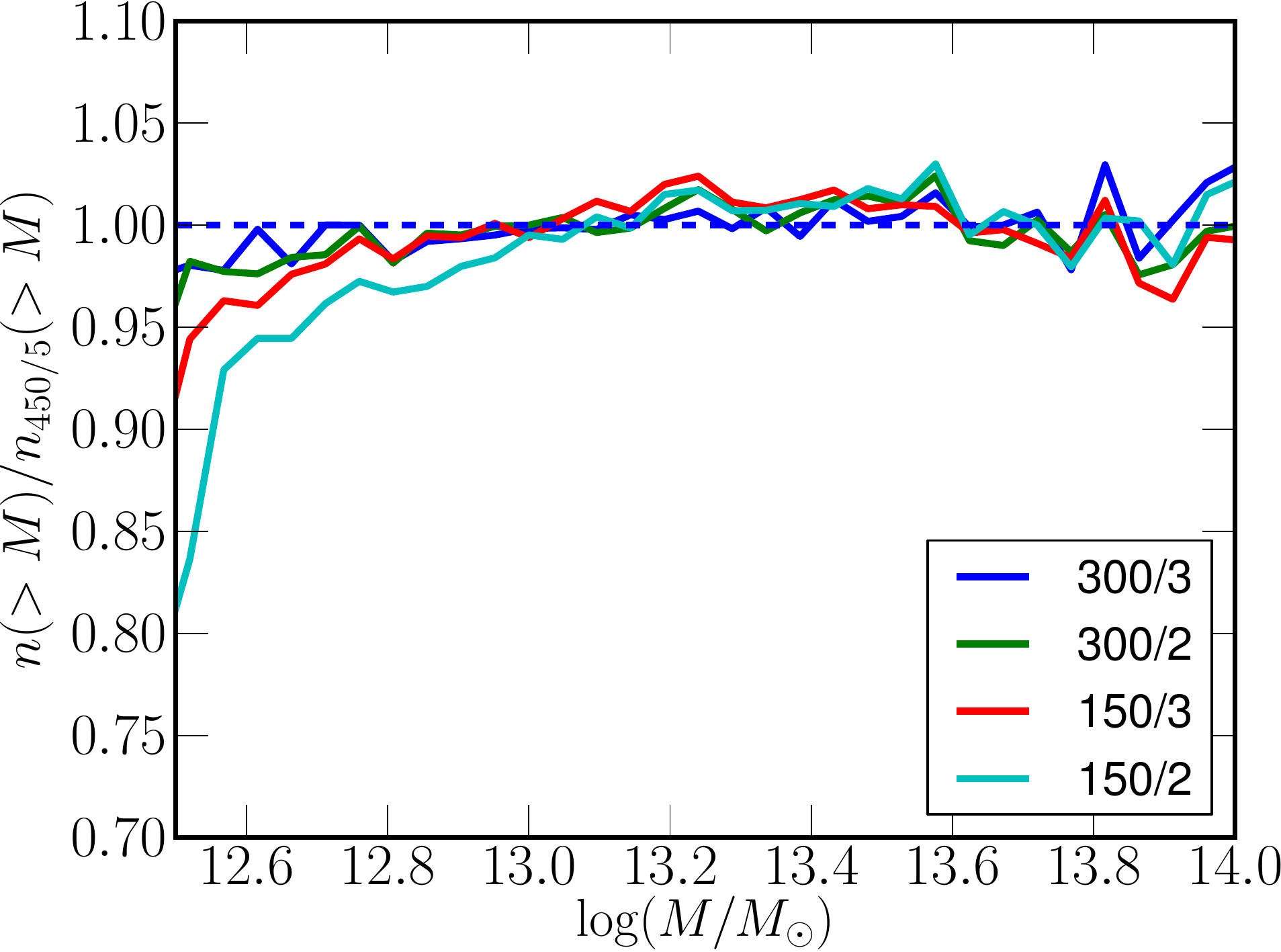} 

\caption{\label{fig:massFn_step}Comparison of cumulative mass
  functions in different simulations taking the 450/5 simulation as a
  reference.  Lines, from top to bottom, correspond to the time
  stepping choices, 300/3 (blue), 300/2 (green), 150/3 (red), and
  150/2 (cyan) respectively.  The left panel shows the cumulative mass
  functions for unadjusted masses (as described in the text), while
  the right panel shows the post-correction results. The simple mass
  recalibration allows us to successfully recover the mass functions,
  even in the extreme case of the 150/2 simulation, for which the
  original result differed by more than 10\% (on all mass scales). }
\end{figure}

Given the mass corrections, we now compute mass functions and halo power spectra using the
results from the different time-stepping schemes. Note that here
we apply the mass adjustment to all the halos in the simulations and use them to compute 
the mass functions and power spectra. As shown in
Figure~\ref{fig:massFn_step}, we use the 450/5 simulation at
$z=0.15$ as the reference. In Figure \ref{fig:massFn_step}, we show
the ratio $n(>M)/n_{450/5}(>M)$, where $n_{450/5}(>M)$ is the
cumulative mass function for the 450/5 simulation and $n(>M)$ is the
cumulative mass function for the other cases. We compare the results
before and after mass adjustment. While the mass functions from the
150/3 and 150/2 simulations are suppressed by more than 10\% on all
mass ranges before correction, they are significantly improved
afterwards, especially for halo masses greater than $10^{13.0}h^{-1}{\rm
  M_{\odot}}$.  For simulations with 300 global time steps, the mass
adjustment is especially effective at small halo masses, and we recover halo masses
to better than 0.5\%.

\begin{figure}[H]
\includegraphics[width=0.45\columnwidth]
{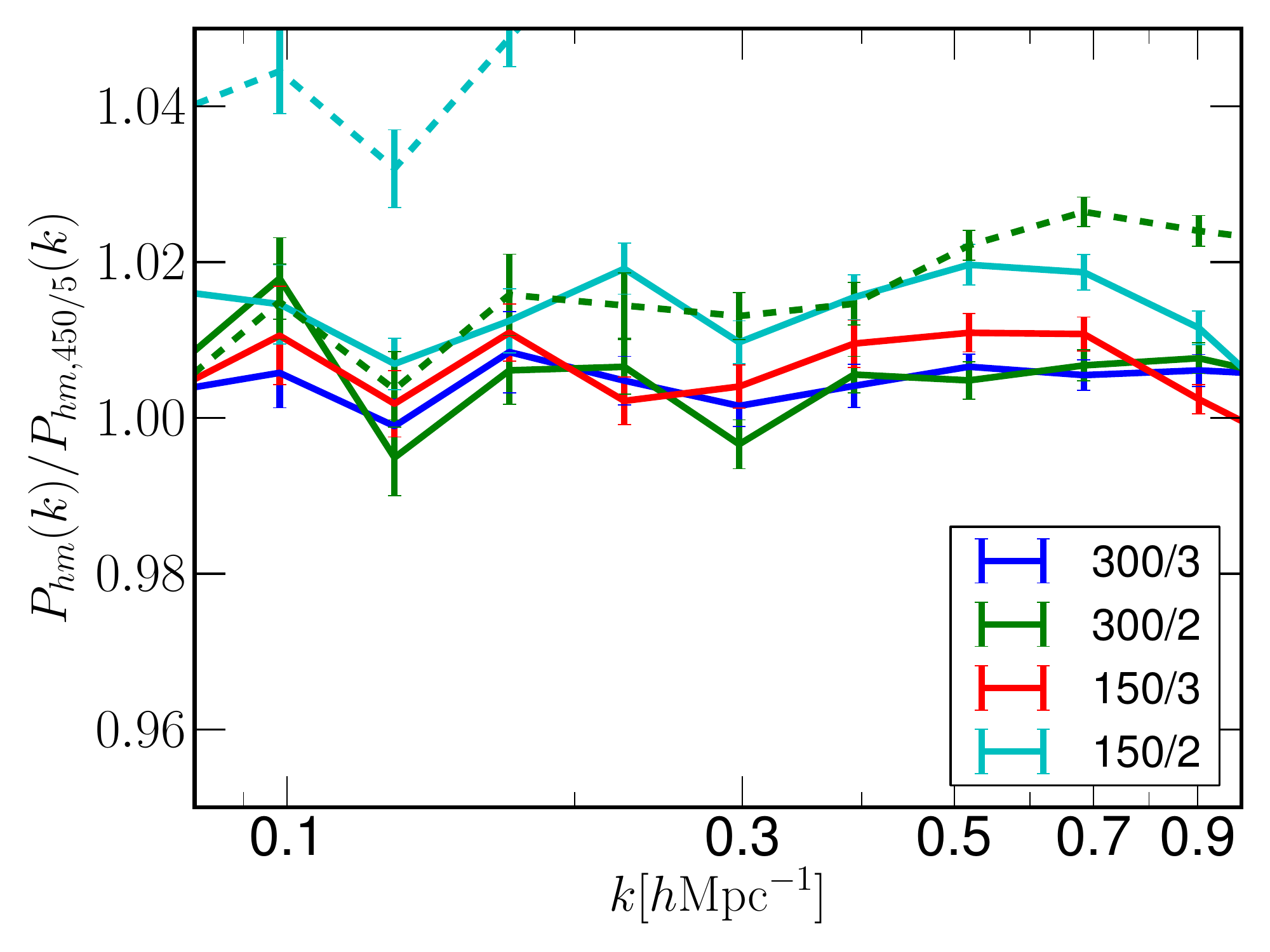}
\includegraphics[width=0.45\columnwidth]
{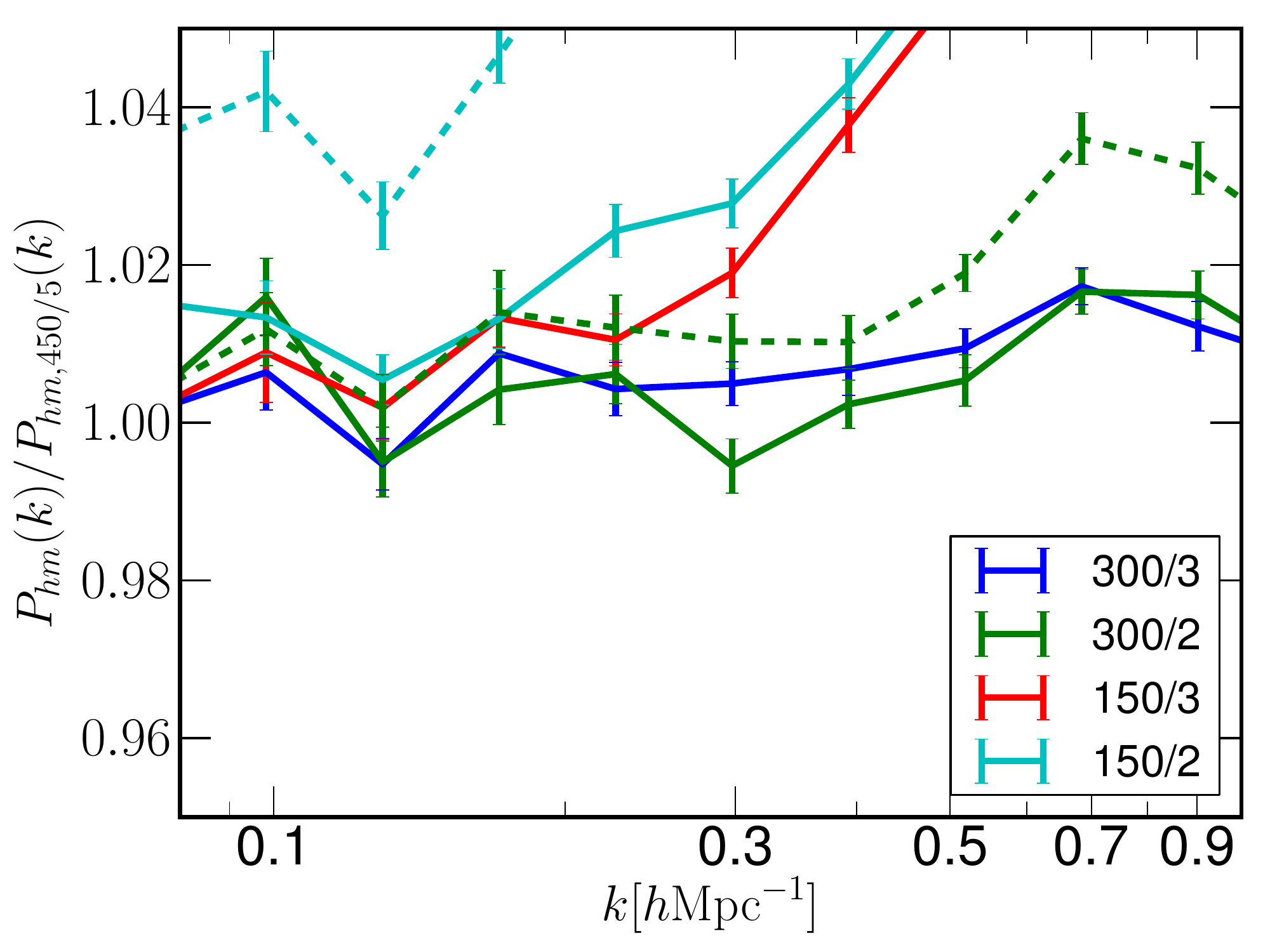}

\caption{\label{fig:crossMater_step} Ratio of halo-matter cross power
  spectra as a function of time steps with respect to the 450/5
  simulation at $z=0.15$. We use the real space halo density field for
  the left panel and the redshift space halo density field for the
  right panel; the dark matter density fields used here are in real
  space for both cases.  The left panel shows that agreements with the
  450/5 simulation are all within 2\%. The dashed lines show the case
  without mass corrections for the 300/2 (green) and the 150/2 (cyan) simulations. In the right
  panel, the large discrepancy of the cross power spectra for the
  simulations with 150 global steps on small scales is mainly due to
  the systematically small velocities shown in Figure
  \ref{fig:HaloProperty_step}. Note that the halos are selected based
  on the soft mass-cut method described in text with ${\rm M_{cut}}=10^{13.0}h^{-1}$M$_\odot$
  and $\sigma=0.5$.}
\end{figure}

Next we compute the halo-matter cross power spectra between halo and
matter density fields in both real and redshift space, as shown in
Figure~\ref{fig:crossMater_step}. This figure shows the ratio
$P_{hm}/P_{hm,450/5}$ at $z=0.15$, where $P_{hm,450/5}$ is the cross
power spectrum for the 450/5 simulation and $P_{hm}$ is the cross
power spectrum for other time steps. For the dark matter density
field, we use the output of the 450/5 simulation for all the halo
samples. Note that the dark matter density fields are given in real
space for both cases. In this way, the ratio $P_{hm}/P_{hm,450/5}$ in
real space is equivalent to the ratio of halo bias between the 450/5
simulation and the simulations with other time-steps. To select halos,
we apply the soft mass cut method using the probability given by
\begin{equation}
\langle N_{halo}(M)\rangle=\frac{1}{2}{\rm erfc}\left(\frac{{\rm
      log_{10}(M_{cut}/}M)}{\sqrt{2}\sigma}\right), 
\end{equation}
where we set ${\rm M_{cut}=10^{13.0}{\it h}^{-1}{\rm M_{\odot}}}$ and
$\sigma=0.5$.  This probability has a similar form to the HOD
technique so that the probability gradually becomes one as halo mass
increases. We use this method to avoid noise from halos scattering
across sharp halo mass boundaries. The errors calculated here are not
due to sample variance as we generate 10 samples from one full sample
with the soft mass cut method. The results show that as the time
stepping is coarsened, the ratio of the cross power spectra increases,
especially in redshift space, where we observe large deviations from
unity on small scales for the 150/2 and 150/3 simulations. This is due
to the overall smaller halo velocities for those simulations, as shown
in Figure~\ref{fig:HaloProperty_step}.  For the simulations with the
300 global time steps, overall agreement with the 450/5 simulation is
almost at the 1\% level on any scale in both real space and
redshift-space. Based on these convergence tests, we conclude that the
300/2 option meets the error requirements.

\section{Constructing Light Cones}
\label{sec:light cones}

One can choose to construct mock catalogs either from a single static
snapshot in redshift or to take the light cone evolution of the halo
and galaxy distribution into account. There are two broad approaches
to the latter problem. The first is to incrementally build up the
light cone while the simulation is running, while the second stitches
static snapshots together at different redshifts to construct an
approximate light cone. Both methods have advantages and
disadvantages. Building the light cone \textquotedblleft{}on the
fly\textquotedblright{} leads to the correct redshift evolution by
construction, but requires simulation post-processing codes (like halo
finders) to be run concurrently with the simulation, making it
nontrivial to change input parameters to these analyses. Using static
time snapshots can be more flexible, but requires a relatively dense
sampling of timesteps. We take this latter approach here; this section
describes how we interpolate between different snapshots and the
requirements on the time step sampling. We note that these results are
more broadly applicable than just for the simulation suite considered
here.

\begin{figure}[H]
\includegraphics[width=0.5\columnwidth]
{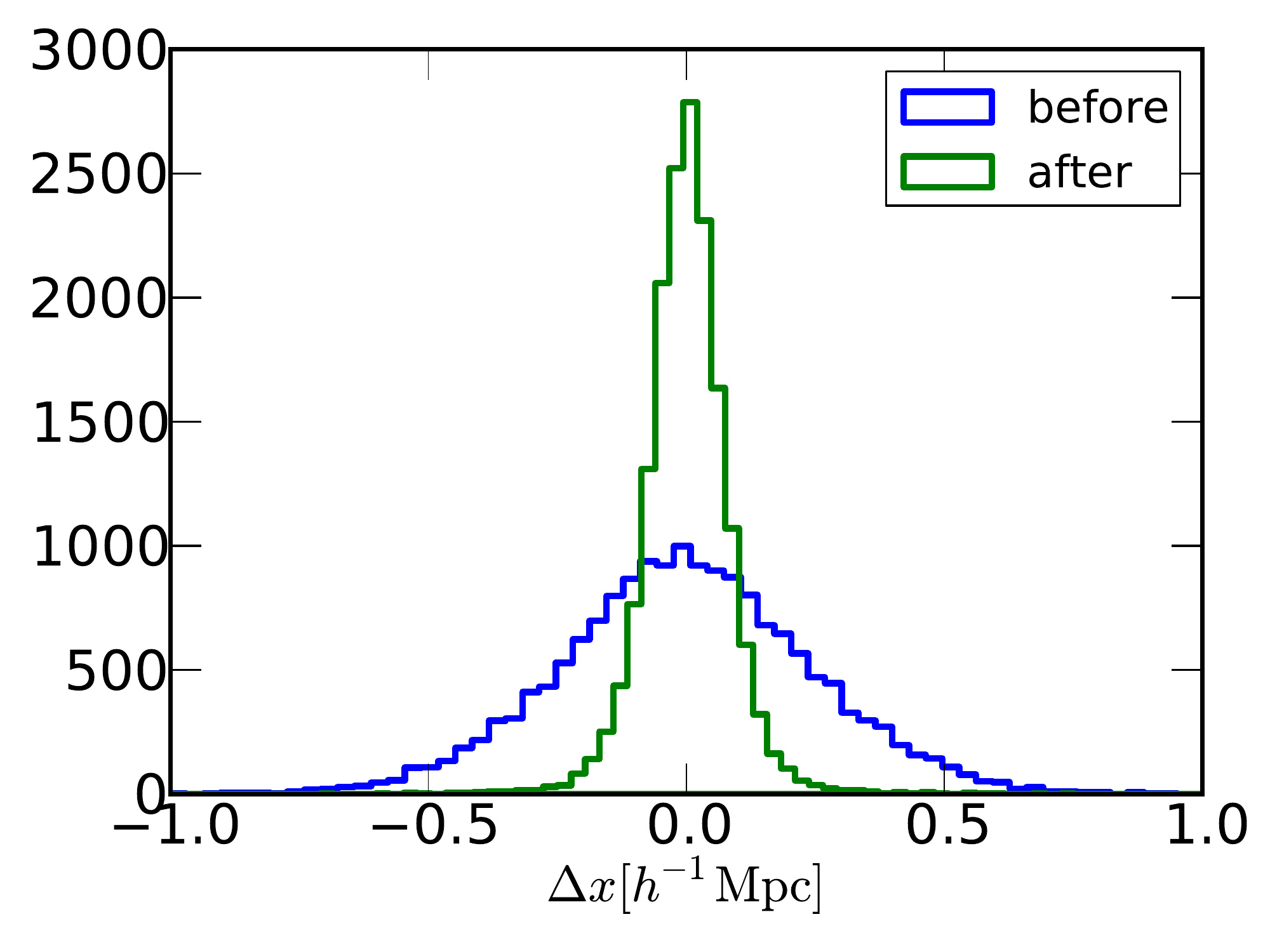}
\includegraphics[width=0.5\columnwidth]
{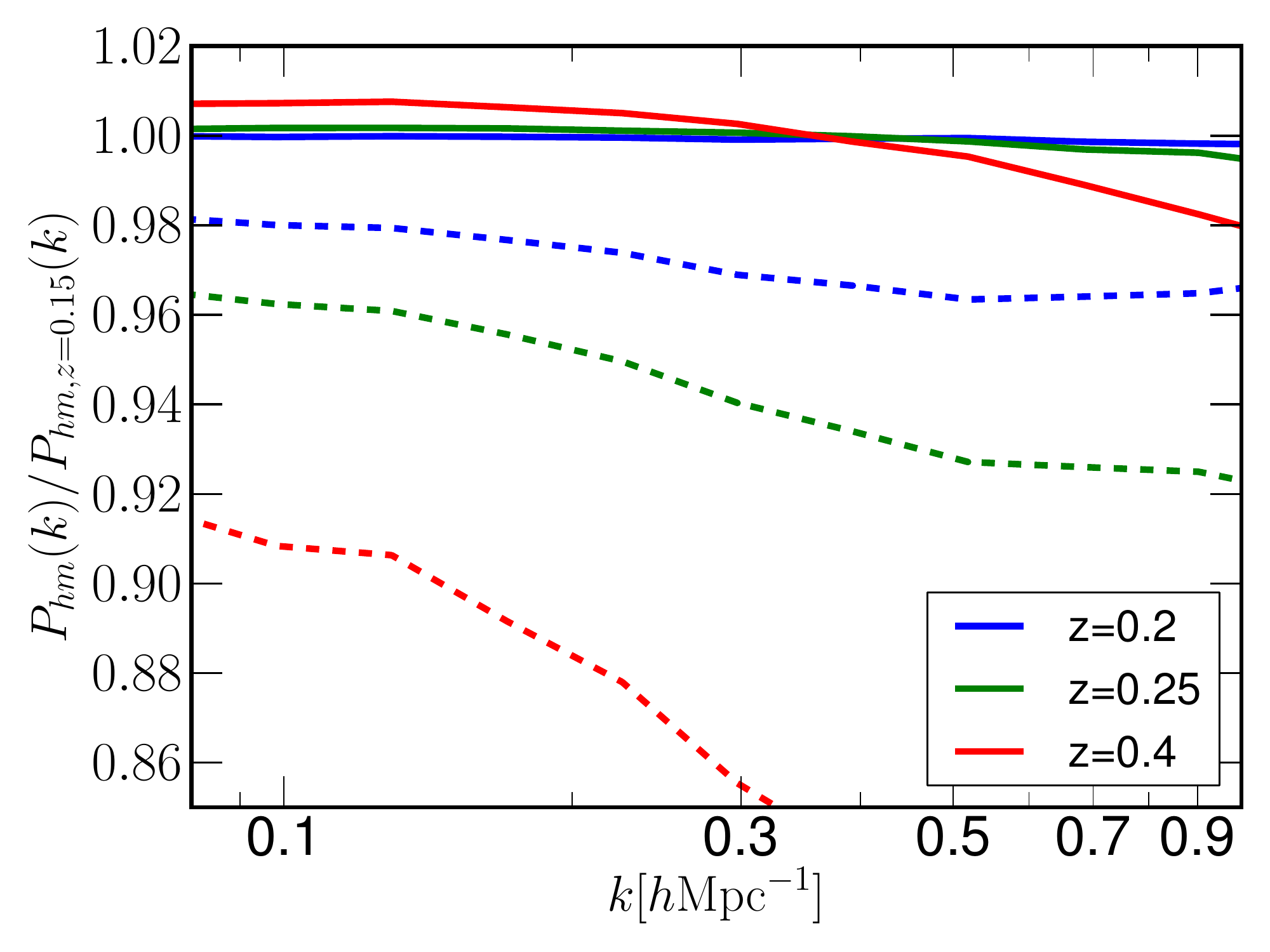}

\caption{Left: Comparison of the $x$-component of the distance before
  and after shifting the positions of halos from $z=0.25$ to
  $z=0.15$. We assume that the peculiar velocity is constant between
  the snapshots (described in Eq. \ref{eq:light cone}) to compute the
  distances to shift. The plot shows that the scatter shrinks to half
  of the original scatter after the halos are shifted. Right: The
  ratio of the power spectrum of the shifted halos to its expected
  value. We shift halos from the redshift shown in the plot ($z=0.4,$
  $0.25$, and $0.2$) to $z=0.15$ and compute the halo-matter cross
  power spectra with the matter density field at $z=0.15$.  The
  denominator is the cross power spectra at $z=0.15$, which is the
  expected value after shifting. The solid line shows the results
  after shifting, while the dashed lines are before shifting. We
  recover the expected power spectrum to better than $1\%$ out to
  $k=1h{\rm Mpc}^{-1}$ for the case of $\Delta
  z<0.25$.}
\label{fig:light cone_pos}
\end{figure}

Our algorithm for constructing a light cone is:
\begin{enumerate}
\item For every snapshot, construct a spherical shell (or part
  thereof) with a radius centered on the comoving distance to the
  snapshot redshift and a width equal to its redshift extent.
\item The redshifts of each halo are determined by their radial
  distance from the origin.
\item Interpolate the position and velocity (see below) of the halo
  from the snapshot redshift to the halo redshift. Note that changing
  the position of the halo strictly changes its redshift, but this is
  a small effect (mostly $<0.1\%$).
\item For the case of a halo crossing its shell boundary, we choose
  the halo whose distance from the boundary is closer before shifting.
\end{enumerate}
We interpolate the positions of halos using a simple linear rule: 
\begin{equation}
\vec{x}|_{z=z_{pos}}=\vec{x}|_{z=z_{snap}}+\vec{v}_{pec}|_{z=z_{snap}}
\Delta t,
\label{eq:light cone}
\end{equation}
where $z_{snap}$ is the redshift of the snapshot, $z_{pos}$ is the
redshift corresponding to the halo's radial position, $\vec{v}_{pec}$
is its peculiar velocity, and $\Delta t$ is the time elapsed between
$z_{snap}$ and $z_{pos}$. We test this using the calibration
simulations discussed in the previous section, restricting ourselves
to halos identified across two or more timesteps. The left panel of
Figure~\ref{fig:light cone_pos} plots the $x$-component of the distance
between halos identified in the $z=0.25$ and $z=0.15$ simulations,
both before and after shifting halos from $z=0.25$ to $z=0.15$ using
the above equation.  The improvement is clearly visible; the scatter
after shifting the halos reduces to half of the original scatter with
similar results for the other components and different redshift
slices. The right panel of the same figure shows the ratio of the
power spectrum of the shifted halos to its expected value. We consider
shifts in redshift ranging from $\Delta z=0.05$ to $\Delta
z=0.25$. Except for the $\Delta z=0.25$ case at high $k$, we recover
the expected power spectrum to better than $1\%$ out to $k=1h{\rm
  Mpc}^{-1}$; the accuracy is a few tenths of a percent for most cases
over most of that range. We obtain similar results for shifting to
different redshifts.

\begin{figure}[H]
\includegraphics[width=0.5\columnwidth]
{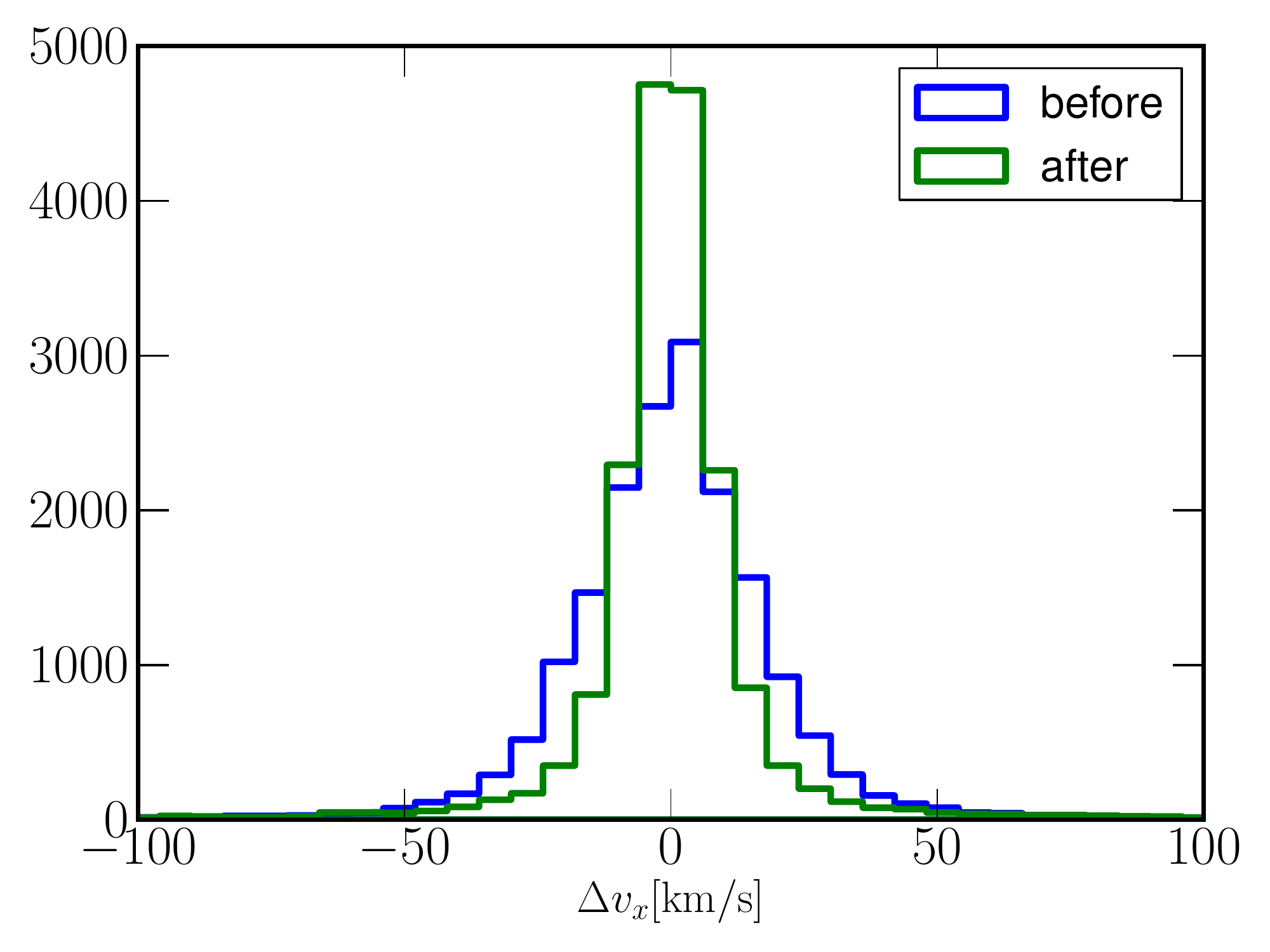}
\includegraphics[width=0.5\columnwidth]
{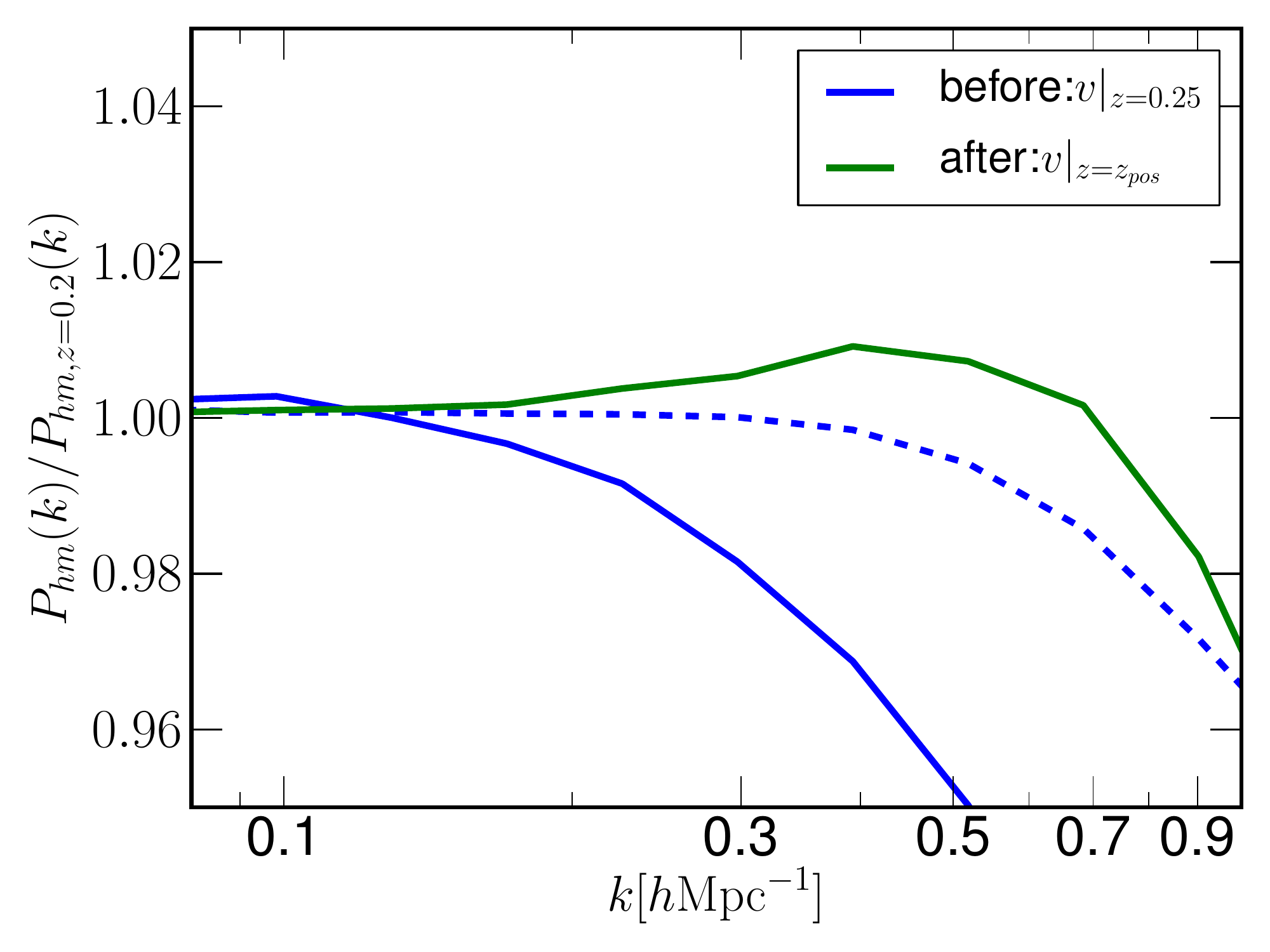}

\caption{\label{fig:light cone_vel}{Left: Comparison of the velocity
    component $v_x$ at $z=0.2$ before and after linear
    interpolation. We estimate velocities at $z=0.2$ by linearly
    interpolating velocities at $z=0.25$ and $z=0.15$. The blue line
    shows the original scatter of the velocities between $z=0.25$ and
    $z=0.2$, while the green line shows the scatter after linear
    interpolation. Right: The ratio of the power spectrum of the
    shifted halos to its expected value in redshift space. We shift
    halos from $z=0.25$ to $z=0.2$ and use original velocities at
    $z=0.25$ (blue line) and the linearly interpolated velocities
    (green line) to define redshift space. The denominator is the
    redshift space cross power spectrum at $z=0.2$, which is the
    expected value after shifting.  As a reference, the result of
    shifting from $z=0.25$ to $z=0.2$ in real space is shown as a
    dashed line. We recover the expected power spectrum to better than
    $1\%$ out to $k=0.8h{\rm Mpc}^{-1}$ by linearly interpolating
    velocities.}}
\end{figure}

Working in redshift space requires interpolating both the positions and the velocities of halos. Since we do not store accelerations of halos, we simply linearly interpolate the halo velocities between two redshifts, $z_{1}$ and $z_{2}$: 
\begin{equation}
\vec{v}|_{z=z_{pos}}=\vec{v}|_{z=z_{1}}\frac{z_{pos}-z_{2}}{z_{1}-z_{2}}+
\vec{v}|_{z=z_{2}}\frac{z_{pos}-z_{1}}{z_{2}-z_{1}}, 
\end{equation}
where $z_{1}<z_{pos}<z_{2}$ and $z_{pos}$ is
the corresponding redshift for positions. Figure~\ref{fig:light cone_vel}
is based on the redshift slices at $z=0.15$ and $0.25$ to predict the values at $z=0.2$; we recover the true velocities with a scatter of 38.8km/s, which is improved from the original scatter of 57.4km/s as shown in the left panel. The right panel of the same figure shows the ratio of the predicted angle-averaged power spectrum to the expected value. While the accuracy is degraded compared to the real space case, we still recover the power spectrum to 1\% out to $k=0.8h{\rm Mpc}^{-1}$. The fall-off at large $k$ is expected since the errors in the velocities introduce a random scatter in the positions of the galaxies, washing out the signal on small scales. 

These results suggest using a redshift spacing of $\Delta z=0.1$ or
better between different simulation outputs. For the simulation suite
discussed in the next section, we choose to be conservative and store
outputs every $\Delta z=0.05$ over the region on interest,
corresponding to a maximum shift in redshift of $0.025$ (smaller than
all the cases considered here).

\section{Example Applications}
\label{sec:boss}

As a concrete implementation of the approach discussed above, we
construct catalogs designed to mimic the Baryon Oscillation
Spectroscopic Survey (hereafter, BOSS) galaxy samples. BOSS~\cite{2013AJ....145...10D}, part of the SDSS-III project~\cite{2011AJ....142...72E}, is a spectroscopic survey that has
achieved percent level distance measurements using the baryon acoustic
oscillation technique~\cite{2014MNRAS.441...24A}.  The low redshift
($z<0.7$) distance measurements use two galaxy samples: the LOWZ
($z<0.45$) and CMASS ($z<0.7$) samples~\cite{2014MNRAS.441...24A,2014MNRAS.440.2222T}.  We describe the
construction of the CMASS sample below. The construction of the LOWZ
sample is analogous, and the same simulations (at different time
slices) can be used in its construction.

We choose a simulation volume large enough to build a full-sky mock
catalog. Since the CMASS sample extends to $z\sim0.7$, we choose a
simulation side of $4000h^{-1}{\rm Mpc}$, corresponding to a comoving
distance to $z\sim0.8$ from the center of the box. We start the
simulations at $z=200$ using Zel\textquoteright{}dovich initial
conditions and evolve them to $z=0.15$ using the time-stepping
procedure described earlier in the paper. We store outputs every
$\Delta z\sim0.05$ starting at $z\sim0.75$; in addition, we store
outputs uniformly spaced by $\Delta z=0.2$ between $z=1$ and $z=2$ as
well as at $z\sim2.5$, $3$ and $4$. The close spacing at low redshift
enables us to make light cones using the method described in the
previous section. We run a friends-of-friends halo finder on each of
the outputs with a linking length of $b=0.168$, keeping halos down to
40 particles, or $10^{12.6}h^{-1}$M$_{\odot}$. By comparison, the
characteristic halo mass of BOSS galaxies is
$10^{13}h^{-1}$M$_{\odot}$, which we resolve with 100 particles. Each
output contains central positions, mean velocities and halo masses,
and $1\%$ of halo particles with a minimum of 5 particles per halo and
$1\%$ of all particles randomly sampled.



\subsection{BOSS Mock Catalogs}

\begin{figure}[H]
\includegraphics[width=0.5\columnwidth]
{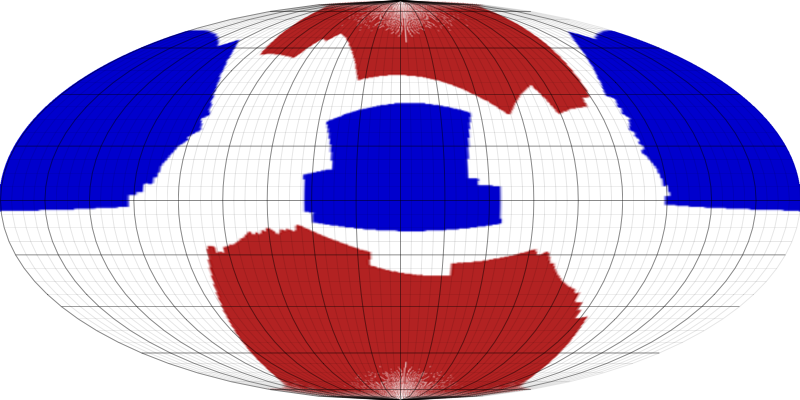}
\label{fig:boss-geom}
\caption{Fitting two non-overlapping BOSS volumes
  into the same simulation box. The blue region is the BOSS survey
  footprint in equatorial coordinates, while the red region is the
  same region rotated using the rotation matrix given in the text.}
\end{figure}

The BOSS angular geometry is split into two regions: one in the North
Galactic Cap and one in the South Galactic Cap
(Figure~\ref{fig:boss-geom}).  Since we generate full-sky mock
catalogs, it is straightforward to embed two full non-overlapping BOSS
surveys in a single mock realization. We cut out a first BOSS volume
with $\vec{x}_{old}$ and then define a new coordinate system
$\vec{x}_{new}$ such that $\vec{x}_{new}=R\vec{x}_{old}$, where $R$ is
the Euler rotation matrix:
\[
R=\left(\begin{array}{ccc}
0.088 & 0.096 & 0.991\\
0.219 & -0.973 & 0.075\\
0.972 & 0.211 & -0.107
\end{array}\right).
\]

To generate the galaxy mock catalogs, we first populate halos with
galaxies using an HOD approach. The HOD functional form (based on a
number of free parameters, 5 in our case) provides probabilities for
the number of central and satellite galaxies based on the masses of
halos that host those galaxies. A halo hosts a central galaxy with
probability $\langle N_{cen}(M)\rangle$ and a number of satellite
galaxies given by a Poisson distribution with mean $\langle
N_{sat}(M)\rangle$:

\begin{equation}
\langle N_{cen}(M)\rangle=\frac{1}{2}{\rm erfc}\left[\frac{{\rm
      ln}(M_{cut}/M)}{\sqrt{2}\sigma}\right],
\label{eq:Ncen}
\end{equation}
and 
\begin{equation}
\langle N_{sat}(M)\rangle=N_{cen}(M)\left(\frac{M-\kappa
    M_{cut}}{M_{1}}\right)^{\alpha},
\label{eq:Nsat}
\end{equation}
where $M_{cut}$, $M_{1}$, $\sigma$, $\kappa$, and $\alpha$ are free
parameters and $M$ is the halo mass. We assume that $N_{sat}(M)$ is
zero when $M<\kappa M_{cut}$ and halos do not host satellite galaxies
without a central galaxy~\cite{2005ApJ...633..791Z}. The total number
of galaxies hosted by each halo is a sum of the number of central and
satellite galaxies. Equations~\ref{eq:Ncen} and \ref{eq:Nsat} are not
the only possible functional form for the HOD, and it is trivial to
change this. However, these forms are known to successfully reproduce
the clustering of the BOSS galaxies~\cite{2011ApJ...728..126W} and are
therefore a convenient choice.

After assigning a number of galaxies to each halo, we distribute those
galaxies within the halo. The central galaxy is always at the center
of the halo, while the distribution of satellite galaxies follows a
spherically symmetric NFW profile specified by: 
\begin{equation}
\rho(r)=\frac{4\rho_{s}}{\frac{cr}{R_{{\rm vir}}}(1+\frac{cr}{R_{{\rm vir}}})^{2}},
\end{equation}
where $\rho_{s}$ is the density at the characteristic scale
$r_{s}=R_{{\rm vir}}/c$ , $R_{{\rm vir}}$ is the virial radius for the
halo and $c$ is the concentration parameter. We use the emulator described in
Ref.~\cite{2013ApJ...768..123K} to generate a table of
concentration-mass relations for halos at each redshift.

We set the velocity of the central galaxy to be equal to the host halo
velocity.  We assume that satellite galaxies are randomly moving
inside the host halos. Therefore, the velocities of the satellite
galaxies are the sum of their host halo velocity and a random virial
component. For this random component, we draw from a Gaussian
distribution with zero mean and variance given by:
\begin{equation}
\langle v_{x}^{2}\rangle=\langle v_{y}^{2}\rangle=\langle
v_{z}^{2}\rangle=\frac{1}{3}\frac{GM}{R_{\rm vir}}.
\label{eq:variance}
\end{equation}
Following the above procedures, we generate two galaxy mock catalogs
from the simulation at $z=0.55$ and from the light cone sample
described in Section~\ref{sec:light cones}. References~\cite{2013MNRAS.428.1036M,2014arXiv1401.4171M}
derived an HOD prescription at $z=0.55$ which was also used in
Ref.~\cite{2014MNRAS.441...24A}. We follow these papers for creating the
mock catalogs here. In addition, we investigate if the mock catalogs
generated from a static time snapshot or from a light cone sample
differ in any significant way.  We use the following HOD parameters,
${\rm log}_{10}M_{cut}=12.9$, ${\rm log}_{10}M_{1}=14.0$, $\alpha=1.013$, $\kappa=1.0$,
$\sigma=0.85$, to populate the halos with galaxies. Note that the goal
here is not to completely fit the observed DR11 correlation functions,
but to show that our simulations provide a reliable way to compute
correlation functions for future galaxy spectroscopic surveys.

\begin{figure}
\includegraphics[width=0.5\columnwidth]
{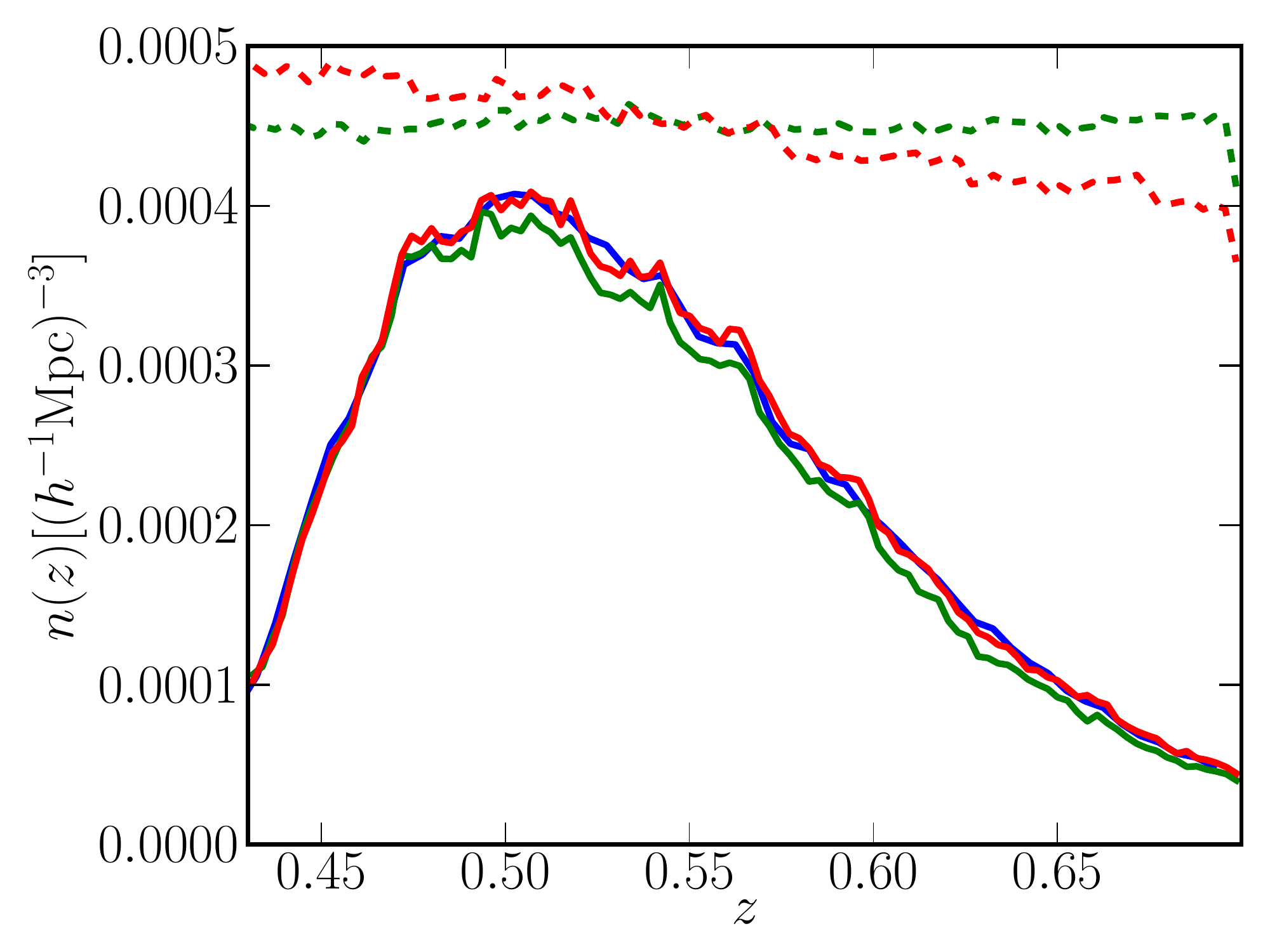}

\caption{\label{fig:nz_gal}Normalized redshift distribution of
  galaxies from DR11 (North) in Ref.~\cite{2013arXiv1303.4666A} (blue
  solid line), and a comparison of galaxy number densities before
  fitting to DR11 redshift selection function (dashed lines) and after
  (solid lines). The green and red lines are from the mock catalogs at
  $z=0.55$ and the light cone output respectively. The HOD parameters
  used to generate the mock catalogs can be found in the text.}
\end{figure}

Once the galaxy mock catalogs are generated, the next step is to match
the number density $n(z)$ to the redshift selection function of
Ref. \cite{2013arXiv1303.4666A}. For each redshift bin, we randomly
subsample galaxies. Figure~\ref{fig:nz_gal} shows redshift
distributions of galaxies before and after subsampling for the BOSS
CMASS North Galactic Cap, which correspond to the dashed and solid lines
respectively. The blue solid line is the redshift distribution of
galaxies for BOSS, while the green and red lines represent $n(z)$ from
the mock catalog at $z=0.55$ and the light cone sample.

\begin{figure}
\includegraphics[width=0.5\columnwidth]{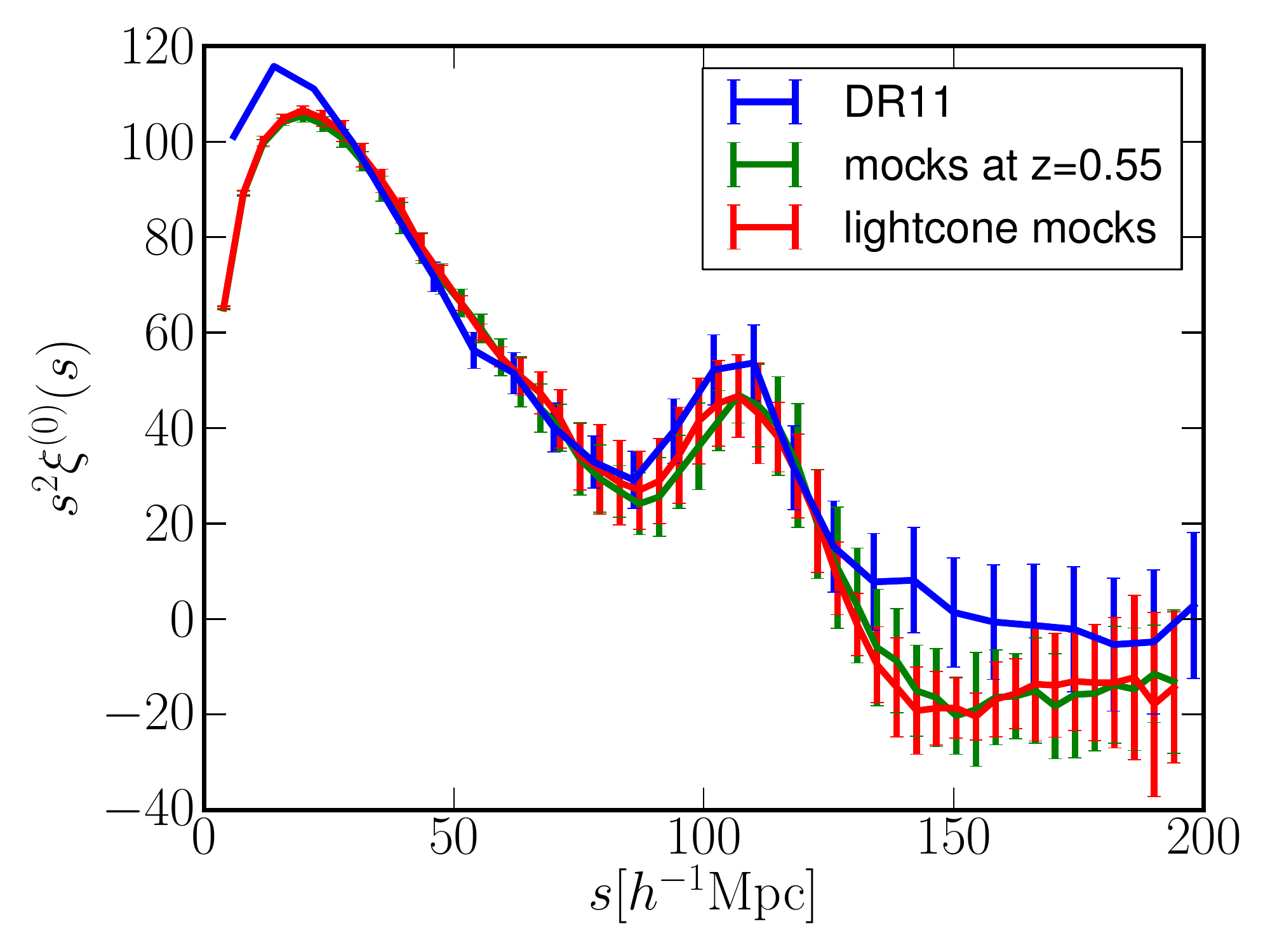}
\includegraphics[width=0.5\columnwidth]
{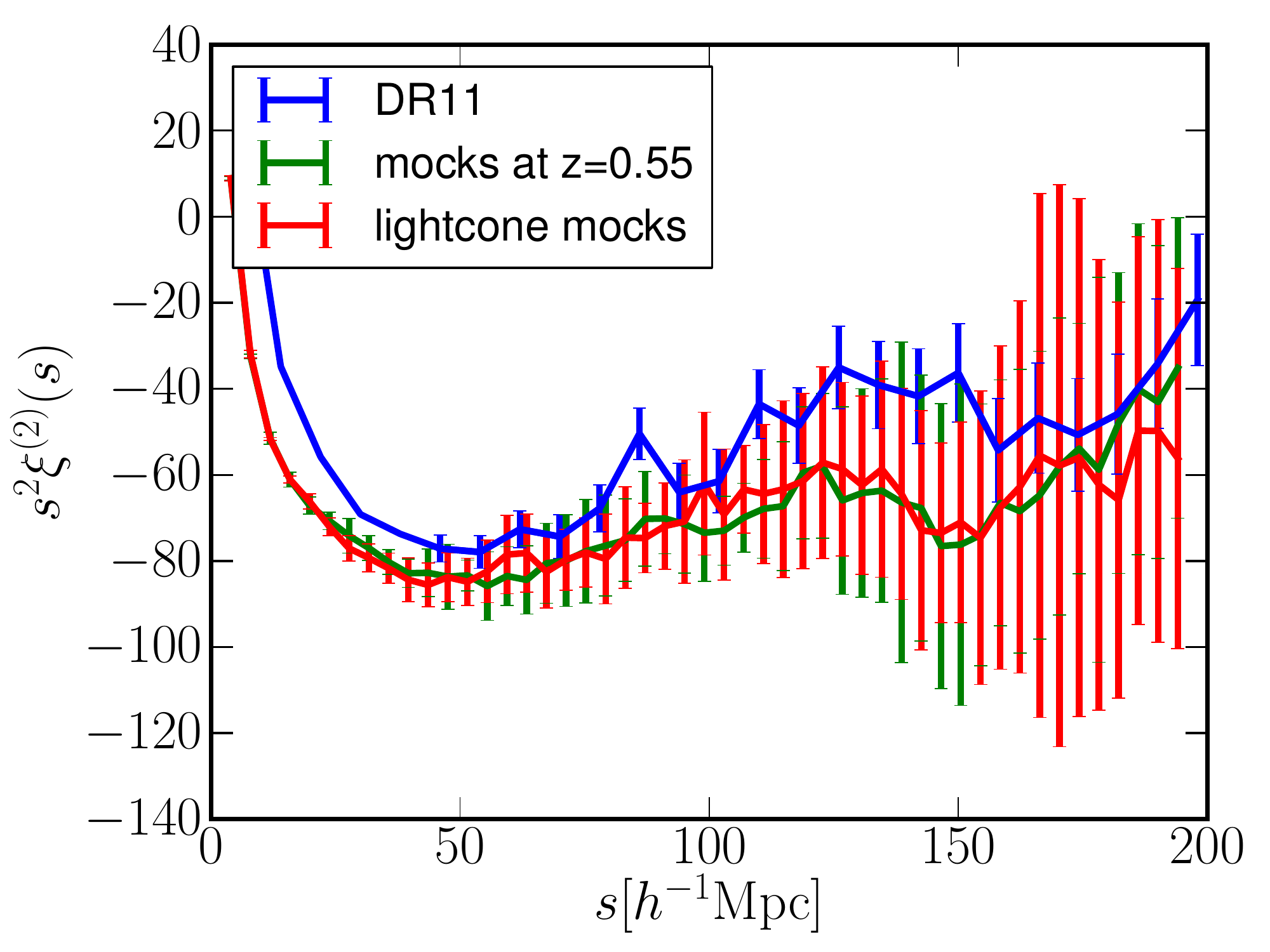}

\caption{\label{fig:xis}Correlation function monopoles $\xi^{(0)}(s)$ (left)
and quadrupoles $\xi^{(2)}(s)$ (right) of the mock catalogs (green and red)
and DR11 in Ref.~\cite{2013MNRAS.428.1036M} (blue) at $z=0.55$.
The HOD parameters used to generate the mock catalogs can be found
in the text.}
\end{figure}

After matching the redshift distribution of galaxies, we finally
compute correlation functions for the galaxy mock catalogs. In
Figure~\ref{fig:xis}, we compare these correlation functions to the
one from DR11. The left panel shows the monopole terms of the
correlation functions, and the right panel, the quadrupole terms. We
do not observe significant differences between the galaxy mock
catalogs from the simulation at $z=0.55$ and the light cone sample for
both monopole and quadrupole terms. For the monopole terms, the mock
catalogs agree relatively well with the one from DR11 on
$r\in[40h^{-1}{\rm Mpc},80h^{-1}{\rm Mpc}]$.  Note that we do not try
to fit to the acoustic peak here, because the cosmologies for our
simulation and DR11 are different. Our prediction for the quadrupole
term differs from the observed quadrupole term from DR11 due to an
overestimate in the power, which is also seen in
Ref.~\cite{2014MNRAS.437.2594W}.

\subsection{Testing the shift of the BAO peaks}
\label{sec:shift}

As a second application for our simulation suite, we explore how the
baryon acoustic oscillation peak changes as a function of galaxy
bias. These shifts, predicted by basic perturbative arguments
(see, e.g., Refs.~\cite{2008PhRvD..77b3533C,2009PhRvD..80f3508P,2012PhRvD..85j3523S}),
must be calibrated at the sub-percent level for future BAO experiments
as a function of the underlying cosmology and galaxy type. Measuring
these shifts requires large simulation volumes to robustly distinguish
them from statistical errors. The approximations we describe in this
paper allow such large volumes to be run, without sacrificing the
gross details (positions, masses, velocities) of the halos themselves
(which, in turn, enables us to construct mock galaxy populations).
Note that our goal here is to demonstrate the statistical power of
these simulations.

In order to construct samples of different galaxy biases, we construct
realizations of HOD models with the same parameters as in Section~5.1,
except for ${\rm log}_{10}M_{cut}$ which we vary from 12.5 to 14.1 in steps of
0.2. The resulting nine HODs span a range of biases from 1.2 to
3.0. In order to estimate the covariance matrix for galaxy correlation
functions, we subdivide each of our $4\times(4h^{-1}{\rm Gpc})^{3}$
volumes into $4\times64=256$ subvolumes. We generate two sets of 
the 256 subboxes by using different random seeds with the same HOD 
parameters to reduce the statistical uncertainties. For simplicity, we analyze
each of these subvolumes individually, except for the ${\rm log}_{10}M_{cut}$=13.9
and 14.1 samples where we analyze groups of 4 subvolumes to reduce the
noise in the measurements.\footnote{In these cases, we scale the
  covariance matrix by a factor of 1/4.} We focus here on real space
measurements. 

We measure the BAO scale using the methodology in
Ref. \cite{2013arXiv1303.4666A}.  Specifically, we describe the
observed correlation function by $\xi_{{\rm fit}}$:
\begin{equation}
\xi_{{\rm fit}}(r)=B\xi_{t}(\alpha r)+A_{0}+A_{1}/r+A_{2}/r^{2},
\label{eq:model1}
\end{equation}
where $\xi_{t}$ is a template correlation function, $B$ is the galaxy
bias squared and $A_{0,1,2}$ represent nuisance parameters to account
for shot noise, nonlinear evolution of the matter density field and
the mapping from matter to galaxies. 
The parameter $\alpha$ measures the shift in the BAO scale relative to
what is assumed for the template function, and is defined by
\begin{equation}
\alpha\equiv\left(\frac{D_{{\rm V}}(z)}{r_{{\rm
        d}}}\right)\left(\frac{r_{{\rm d,fid}}}{D_{{\rm V}}^{{\rm
        fid}}(z)}\right),
\label{eq:alpha_ratio}
\end{equation}
where $r_{{\rm d}}$ is the sound horizon at the
drag epoch and 
\begin{equation}
D_{{\rm V}}\equiv\left[cz(1+z)^{2}D_{{\rm A}}(z)^{2}H^{-1}(z)\right]^{1/3}
\end{equation}
with $D_{{\rm A}}$ being the angular diameter distance and $H(z)$, the
Hubble parameter. In the absence of systematics, we expect $\alpha=1$;
deviations from this represent a bias to the inferred distance-redshift 
relation.

The template correlation function
$\xi_{t}(r)$ is given by a Fourier transform of $P_{t}(k)$:
\begin{equation}
P_{t}(r)=(P_{{\rm lin}}(k)-P_{nw}(k))e^{-\frac{k^{2}\Sigma^{2}}{2}}+P_{nw}(k),
\end{equation}
where $P_{{\rm lin}}(k)$ is the linear power spectrum and $P_{nw}(k)$
is the no-wiggle power spectrum described in
Ref. \cite{1998ApJ...496..605E}, and $\Sigma$ is a nonlinear parameter
that accounts for the broadening of the BAO peak due to nonlinear
evolution. We set $\Sigma=5h^{-1}{\rm Mpc}$.  We determine the
parameters by minimizing $\chi^{2}=(\xi_{{\rm HOD}}-\xi_{{\rm
    fit}})^{{\rm T}}C^{-1}(\xi_{{\rm HOD}}-\xi_{{\rm fit}})$ where
$C^{-1}$ is the inverse covariance matrix and $\xi_{{\rm HOD}}$ is a
correlation function which $\xi_{{\rm fit}}$ is fitted to. We consider
the measured correlation function from $60h^{-1}{\rm Mpc}$ to
$160h^{-1}{\rm Mpc}$ in bins of $4h^{-1}{\rm Mpc}$ for a total of 25
data points. Since all parameters except $\alpha$ are linear, we
perform the minimization on a grid of $\alpha$ values, computing the
minimum value of the other parameters directly. The parameter $\alpha$
measures the shift of the BAO peak from its original position
predicted by linear perturbation theory. Note that $\alpha=1$ implies
that there is no shift of the BAO peak.

\begin{figure}[H]
\includegraphics[width=0.5\columnwidth]{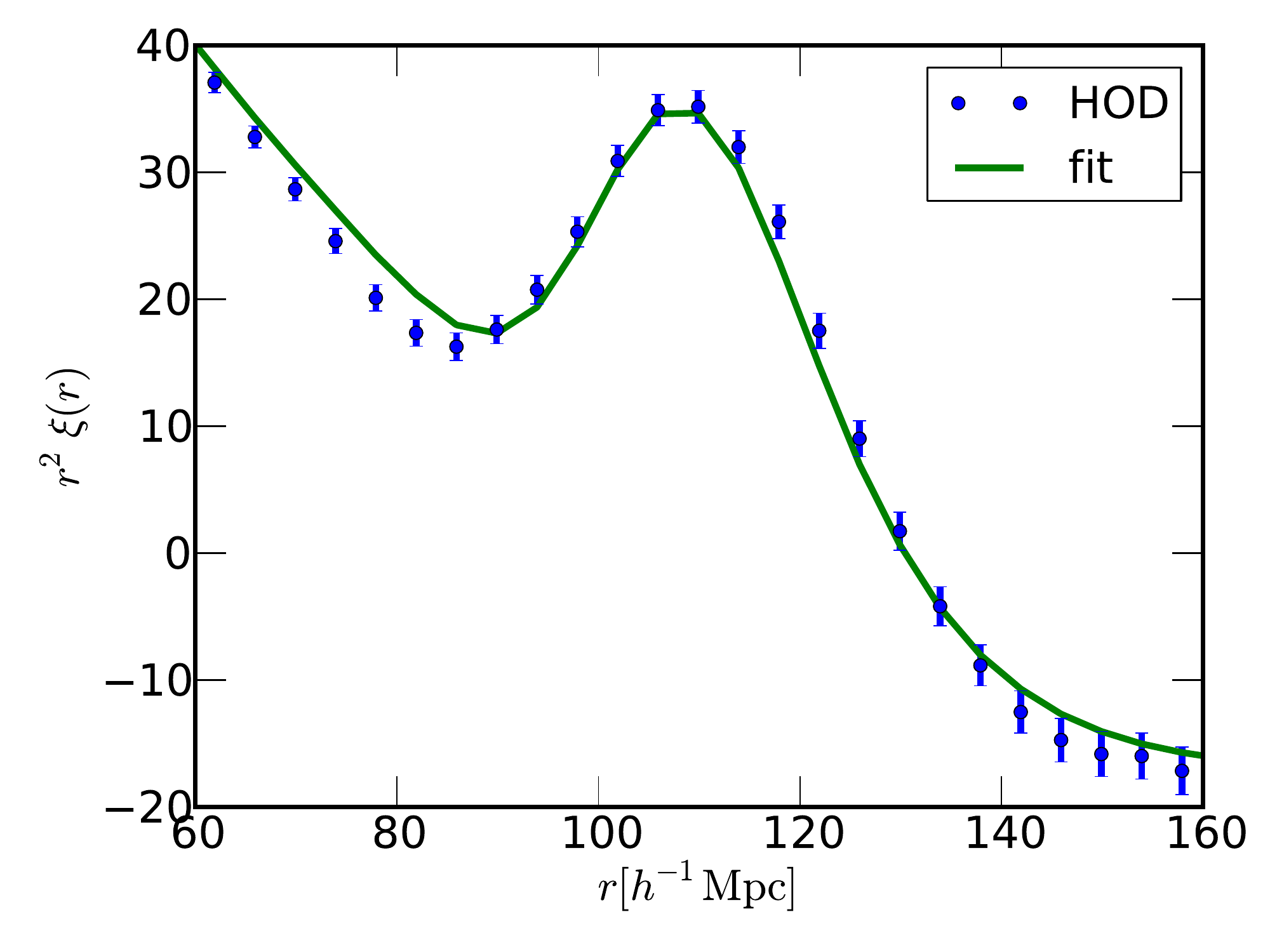}
\includegraphics[width=0.5\columnwidth]
{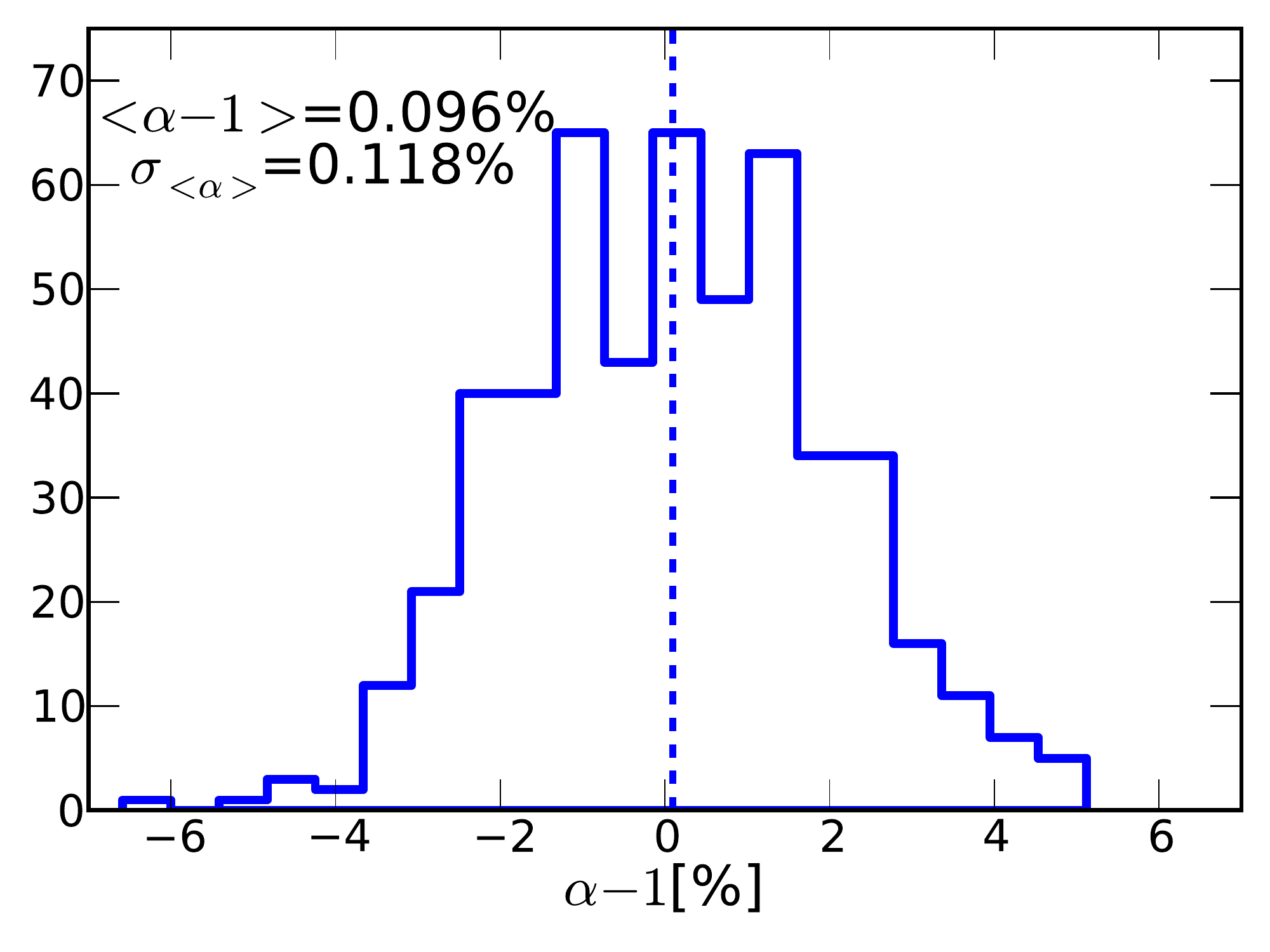}

\caption{\label{fig:demo}Left: The correlation function $\xi_{{\rm
      HOD}}(r)$ computed from the full HOD galaxy mock catalog
  (labeled as ``HOD'') with ${\rm log}_{10}M_{cut}=12.9$ (which corresponds to the
  bias value of 1.43) and the template correlation function with the
  best-fit parameters (labeled as ``fit''). The best fit value of
  $\alpha$ for this mock is 1.005. The error bars are computed from
  the covariance matrix. Right: Distribution of the values $\alpha-1$
  for the case of ${\rm log}_{10}M_{cut}=12.9$.  The dashed line corresponds to the
  mean value of $\alpha-1$, which is 0.1 \%.}
\end{figure}

The left panel of Figure~\ref{fig:demo} compares the correlation
function computed from one of the full boxes with ${\rm log}_{10}M_{cut}=12.9$
(which corresponds to a galaxy bias of 1.43) and the model
correlation function described in Eq.~\ref{eq:model1} with the
best-fit parameters.  The best fit value of $\alpha$ for this mock is
1.005. Note that the error bars shown in the left panel of
Figure~\ref{fig:demo} are computed from the covariance matrix. The
right panel of Figure~\ref{fig:demo} shows the distribution of
$\alpha-1$ for the case of ${\rm log}_{10}M_{cut}=12.9$ for the 256 samples. The
mean value of $\alpha-1$ is $0.1\pm0.1\%$.

\begin{figure}[H]
\includegraphics[width=0.5\columnwidth]{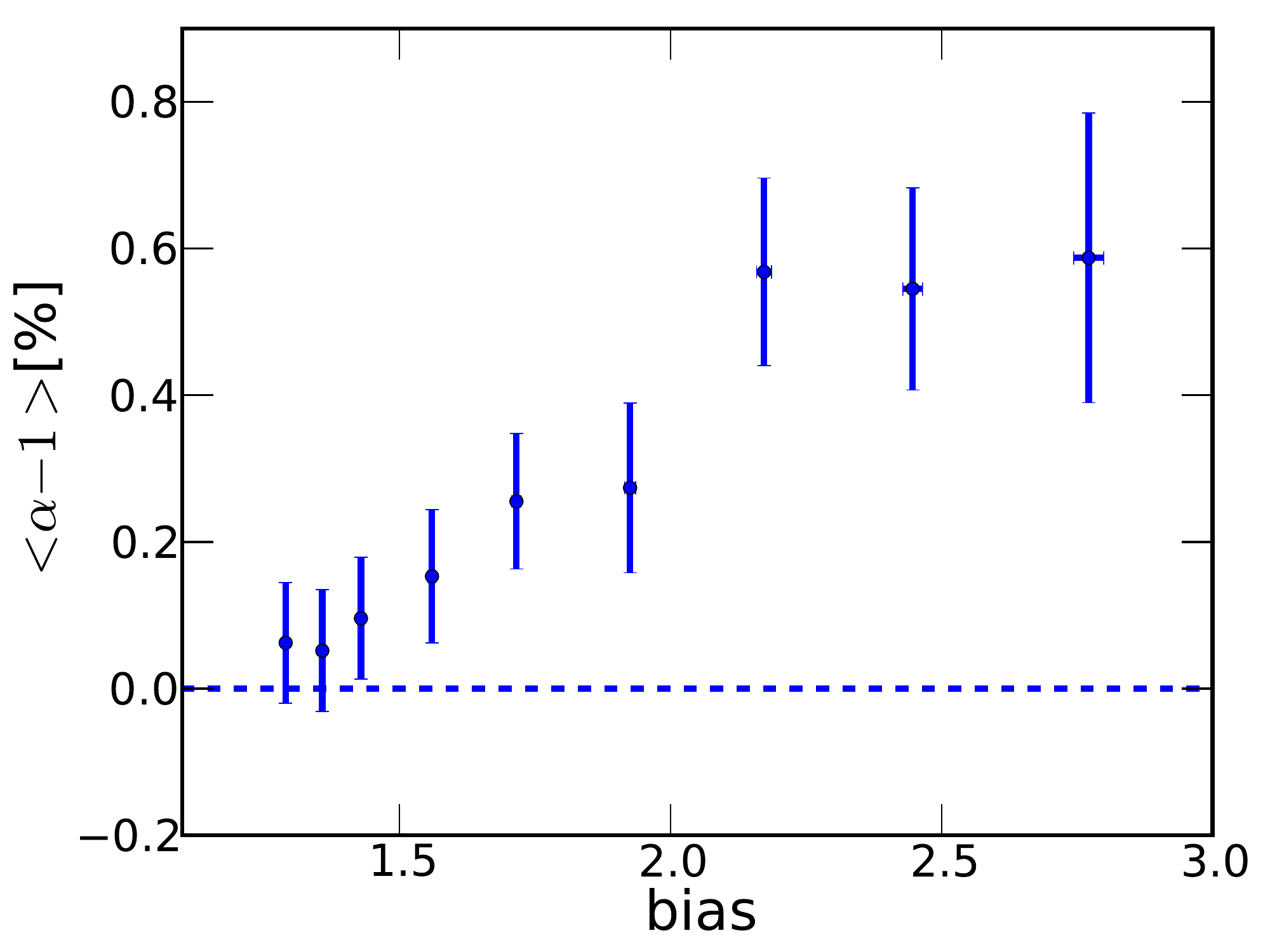}

\caption{\label{fig:alpha}Shift in the BAO scale as a function of
  galaxy bias. Note that $\alpha=1$ implies no shift in the BAO
  scale. We detect a shift and a variation with the galaxy bias. These results are slightly different from the trends shown in 
Ref.~\cite{2011ApJ...734...94M}, which did not find a variation with the galaxy biases
$<$ 3.0. The third point from left is for the HOD parameter value
 ${\rm log}_{10}M_{cut}=12.9$, the same as used in Figure~\ref{fig:demo}. }
\end{figure}

Figure~\ref{fig:alpha} shows the measured shifts in the BAO scale as a function of galaxy 
bias, for the nine HODs described above. We detect the shift in the BAO scale and a variation 
with the galaxy bias. These results are slightly different from the trends shown in 
Ref.~\cite{2011ApJ...734...94M}, which did not find a variation with the galaxy biases
$<$ 3.0, but Ref.~\cite{2011ApJ...734...94M} uses samples with smaller volumes at $z=1$.

We also find that our results show 
a mass dependence roughly consistent with what was predicted in 
Ref.~\cite{2009PhRvD..80f3508P}, although we find an indication of somewhat weaker 
redshift dependence expected from perturbation 
theory~\cite{2008PhRvD..77b3533C,2009PhRvD..80f3508P,2012PhRvD..85j3523S} for 
low biased objects. We defer a detailed comparison with the results from perturbation 
theory~\cite{2008PhRvD..77b3533C,2009PhRvD..80f3508P,2012PhRvD..85j3523S} as 
well as previous simulations for future work; these results highlight the utility of the 
large volumes enabled by these simulations for such studies.

This work can be extended by using reconstruction methods to verify
that it does indeed reduce these biases both in real space and in
redshift space, comparing to perturbation theory results, and testing
redshift evolution and cosmology dependence in the shift of the peak.

As a final example, we use our simulations to compute the 
distance scale to $z=0.55$, using the published DR11 correlation 
functions presented earlier. We do so using the same methodology 
as above, except that we use the average correlation function 
obtained from the simulations to define the template correlation function.
This highlights a different use of these simulations -- to build accurate
models for the galaxy correlation function.

\begin{figure}[H]
\includegraphics[width=1\columnwidth]{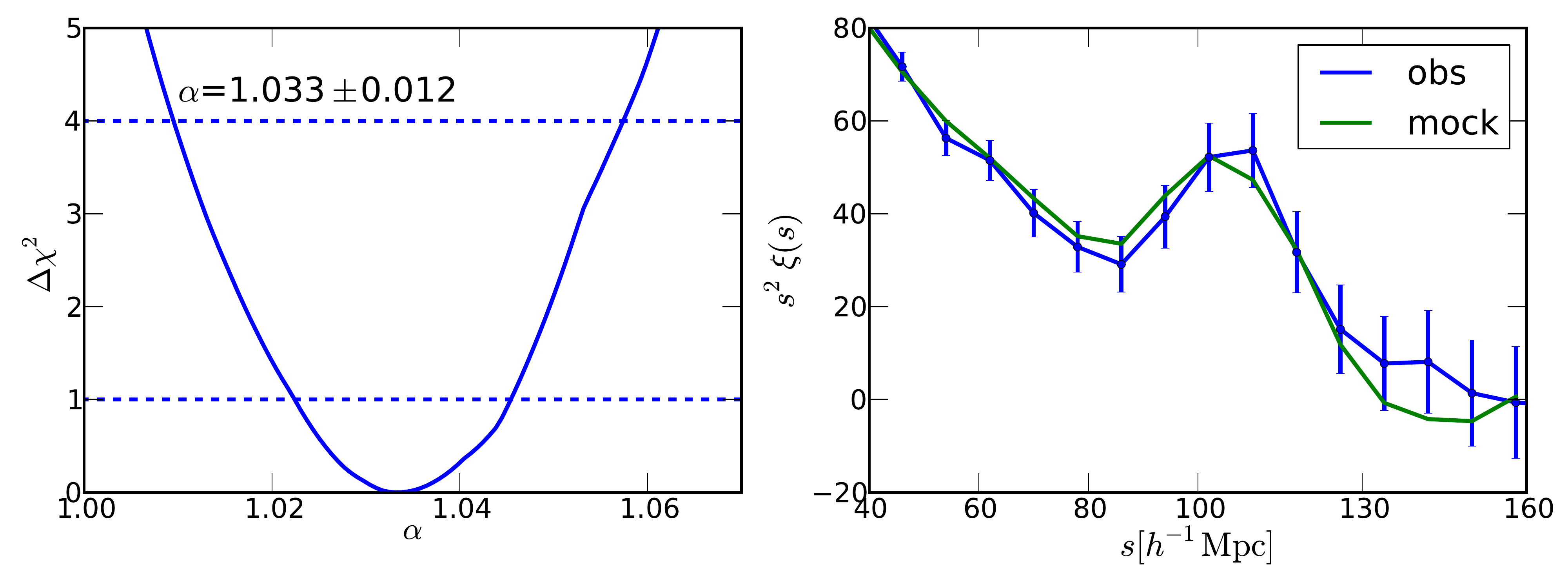}

\caption{\label{fig:alpha_fit}Left: Plot of $\Delta\chi^{2}$
  vs. $\alpha$ for the data from DR11. The dashed lines (from top to
  bottom) correspond to $2\sigma$ and $1\sigma$ for the value of
  $\alpha$. The best fit value of $\alpha$ is $1.033\pm0.012$. Right:
  The monopole correlation functions of the data from DR11 (blue line)
  and of the light cone galaxy mock fitted with $\alpha=1.033$.}
\end{figure}

Figure~\ref{fig:alpha_fit} shows the $\chi^{2}$ values as a function
of the parameter $\alpha$ in the left panel and compares the monopole
correlation function from DR11 and the fitted function in the right
panel. 
The ratio $r_{{\rm d,fid}}/D_{{\rm V}}^{{\rm
    fid}}(z)$ for our simulations is 13.44 at $z=0.57$, while the
ratio $D_{{\rm V}}(z)/r_{{\rm d}}$ from DR11 is
$13.91\pm0.13$. Therefore, the estimated value for $\alpha$ is
$1.025\pm0.010$.
We obtain a value of $1.033\pm0.012$, which is within $1\sigma$
from the estimated value for $\alpha$. 

\section{Discussion}
\label{sec:disc}

The precision measurements expected from current and future galaxy
spectroscopic surveys to test the expansion and structure formation
history of the Universe require an accurate understanding of
systematic effects. To explore these effects, large-volume N-body
simulations with acceptable accuracy are required. De-tuning
high-resolution N-body codes by coarse-graining the time stepping
offers a way to generate these simulations at lower computational
cost. In this paper we have presented a quantitative study of the
impact of time step sizes on the halo and matter density fields using
the HACC code. HACC has two adjustable time stepping parameters -- a
global time step and sub-cycle steps to track particle trajectories on
small scales. We consider cases where we increase the length of each
time step by factors of 1.5 and 3 respectively, as well as reducing
the number of sub-cycles. We find that the number of sub-cycles
makes almost no difference to any of our final results.

Our fiducial choice is to use 300 global time steps corresponding to
$\Delta a(z)=0.003$ and 2 sub-cycles (increasing the length of the
global time step by a factor of 1.5 and that of the sub-cycles by a
factor of 2.5), resulting in a reduction of the simulation run times
by a factor of four. We keep the mass resolution constant; the results
here are based on a particle mass of $6.86\times10^{10}h^{-1}{\rm
  M_{\odot}}$. We summarize the key results below:

(a) Halo masses are underestimated because reducing the number of time
steps produces halos with less substructure and a more diffuse
distribution of mass. We demonstrate that this can be corrected with
higher fidelity simulations, recovering the halo masses to 99.5\%
fidelity. The halo mass function is correctly recovered for
masses above $10^{12.7}h^{-1}{\rm M_{\odot}}$ corresponding to 100
particles per halo. We run the halo finder with identical parameters
as in the full resolution runs for simplicity. It may however be possible to get
further improvements by changing the parameters of the halo finder, as was
done in Ref.~\cite{2013MNRAS.428.1036M}. We leave an exploration of this for 
future work.

(b) The halo positions and velocities are recovered with a scatter of
$0.08h^{-1}{\rm Mpc}$ and $13.4{\rm km/s}$ respectively, ensuring 
that their large scale distributions are unaffected.

(c) The clustering of halos is correctly recovered to better than 1\%
on scales below $k<1h{\rm Mpc^{-1}}$ in real space and $k<0.5h{\rm
  Mpc^{-1}}$ in redshift space.
  
We also consider the frequency with which one must store outputs to
construct galaxy light cones. In real space, a simple linear interpolation
over $\Delta z < 0.25$ recovers the power spectrum to better than 1\%
for $k<1 h {\rm Mpc}^{-1}$. In redshift space, we find we additionally 
need to interpolate halo velocities. Again, linear interpolation 
recovers the redshift space power spectrum to 1\% for $k < 0.8 h {\rm Mpc}^{-1}$.
We conclude that spacings of $\Delta z=0.05$ or $0.1$ are adequate for
constructing light cones from single-time snapshots. We note that these
requirements are significantly less stringent than what is required for 
constructing merger trees, and therefore will often be trivially met.

As a demonstration of the utility of the methods presented here, we 
generate a suite of four $(4 h^{-1}{\rm Gpc})^3$ simulations for a total
simulation volume of 256 $h^{-3} {\rm Gpc}^3$. We consider two example 
applications of these. The first is the construction of mock galaxy catalogs.
We focus here on the BOSS CMASS sample, but these simulations
may be used for future surveys like eBOSS and DESI. The second demonstration
is to measure the effects of nonlinear evolution and galaxy bias on the position
of the BAO feature. This extends the work of Ref.~\cite{2011ApJ...734...94M}
with significantly improved statistics. We find shifts in the BAO peak as a 
function of galaxy bias. Both of these examples
were chosen to demonstrate the utility of these simulations, and will be 
explored in more detail in future work.

The above examples are not the only applications of the simulations
made possible by our approach here. For instance, while the
volumes of simulations are still not large enough to simply measure 
a sample covariance matrix, such simulations could be very useful in 
calibrating models of covariance matrices. Unlike other approximate methods, 
these simulations are accurate down to relatively small scales $k \sim 1 h {\rm Mpc}^{-1}$,
and so are ideal for comparing against models.

Large scale suites of simulations are an essential part of current and future
cosmological analyses. This paper helps lay part of the groundwork for how one can 
optimize the computational requirements of such simulations.

\section*{Acknowledgement}

NP and TS acknowledge support from the DOE Early Career Grant DE-SC0008080. This work was
supported in part by the facilities and staff of the Yale University
Faculty of Arts and Sciences High Performance Computing Center. TS would like to thank Andrew Szymkowiak for useful discussions. SH, KH, and ER record their indebtedness to other HACC team members -- Hal Finkel, Nicholas Frontiere, Vitali Morozov, Adrian Pope -- for their assistance and contributions. The work of SH, KH, and ER at Argonne National Laboratory work was supported
under the U.S. Department of Energy contract DE-AC02-06CH11357.
This research used resources of the National Energy Research Scientific
Computing Center, a DOE Office of Science User Facility supported
by the Office of Science of the U.S. Department of Energy under Contract
No. DE-AC02-05CH11231. 

\bibliographystyle{JHEP}
\bibliography{bigsims}

\end{document}